\documentclass[traditabstract]{aa}

\usepackage{graphicx}
\usepackage{savesym}
\savesymbol{rotate}
\usepackage{deluxetable} 
\restoresymbol{DTBL}{rotate}
\usepackage{times}
\usepackage{epsfig} 
\usepackage{mathrsfs}  
\usepackage{natbib}
\usepackage{amssymb}
\usepackage{amstext}
\bibpunct{(}{)}{;}{a}{}{,} 
\usepackage{caption}
\usepackage[english]{babel}
\usepackage{longtable}
\usepackage{url}
\usepackage{hyperref}
\usepackage{xspace}

\usepackage{color}

\newcounter{aaffilcoun}
\newcommand\theaaffil{\addtocounter{aaffilcoun}{1}\theaaffilcoun}
\newcounter{affilcoun}

\def\ifundefined#1{\expandafter\ifx\csname#1\endcsname\relax}
\ifundefined{ensuremath}\def\ensuremath#1{\relax\ifmmode{#1}}
\else${#1}$\fi\else\relax\fi
\ifundefined{nuc}\def\nuc#1#2{\relax\ifmmode{}^{#1}{\protect\mathrm{#2}}
\else${}^{#1}$#2\fi}\else\relax\fi

\newcommand{\nni}{\ensuremath{\nuc{56}{Ni}}\xspace}
\newcommand{\nco}{\ensuremath{\nuc{56}{Co}}\xspace}

\newcommand{\vphot}{\ensuremath{v_{\text{phot}}}\xspace}

\newcommand{\gcm}{g~cm$^{-3}$\xspace}
\newcommand{\kmps}{\ensuremath{\mathrm{km}~\mathrm{s}^{-1}}\xspace}

\newcommand{\msol}{\ensuremath{{\mathrm{M}_\odot}}\xspace}
\def\la{\mathrel{\hbox{\rlap{\hbox{\lower4pt\hbox{$\sim$}}}\hbox{$<$}}}}
\def\ga{\mathrel{\hbox{\rlap{\hbox{\lower4pt\hbox{$\sim$}}}\hbox{$>$}}}}
\authorrunning{Stritzinger et al.}
\titlerunning{The bright and energetic Type~Iax SN~2012Z.}
\begin{document}

\title{{Comprehensive Observations of the Bright and Energetic Type~Iax
  SN~2012Z: Interpretation as a Chandrasekhar Mass White Dwarf Explosion}
\thanks{Based on
    observations collected at the European Organization for
    Astronomical Research in the Southern Hemisphere, Chile (ESO
    Program 088.D-0222, 184.D-1152), the Magellan 6.5~m telescopes at
    Las Campanas Observatory, and the Nordic Optical Telescope,
    operated by the Nordic Optical Telescope Scientific Association at
    the Observatorio del Roque de los Muchachos, La Palma, Spain, of
    the Instituto de Astrofisica de Canarias.  Also partially based on
    observations made with the Southern African Large Telescope
    (SALT), and the W. M. Keck Observatory located on the summit of
    Mauna Kea.}}

\author{M.~D. Stritzinger\inst{\theaaffil}
\and S. Valenti\inst{\theaaffil,\theaaffil}
\and P. Hoeflich\inst{\theaaffil}
\and E. Baron\inst{\theaaffil,\theaaffil}
\and M. M. Phillips\inst{\theaaffil}
\and F. Taddia\inst{\theaaffil}
\and R. J. Foley\inst{\theaaffil}
\and E. Y. Hsiao\inst{1,7}
\and S. W. Jha\inst{\theaaffil}
\and C. McCully\inst{10}
\and V. Pandya\inst{10} 
\and J. D. Simon\inst{\theaaffil}
\and S. Benetti\inst{\theaaffil}
\and P. J. Brown\inst{\theaaffil}
\and C.~R. Burns\inst{11}
\and A. Campillay\inst{7}
\and C. Contreras\inst{1,7}
\and F. F\"{o}rster\inst{\theaaffil}
\and S. Holmbo\inst{1}
\and G. H. Marion\inst{\theaaffil}
\and N. Morrell\inst{7}
\and G. Pignata\inst{\theaaffil}
}

\institute{Department of Physics and Astronomy, Aarhus University, Ny Munkegade 120, DK-8000 Aarhus C, Denmark\\ (\email{max@phys.au.dk})
\and
Las Cumbres Observatory Global Telescope Network, Inc.
Santa Barbara, CA 93117, USA
\and
Department of Physics, University of California, Santa Barbara, Broida Hall, Mail Code 9530, Santa Barbara, CA 93106-9530, USA
\and
Department of Physics, Florida State University, Tallahassee, FL 32306, USA
\and
Homer L. Dodge Department of Physics and Astronomy, University of Oklahoma, 440 W. Brooks, Rm 100, Norman, OK 73019-2061, USA
\and
Hamburger Sternwarte, Gojenbergsweg 112, 21029 Hamburg, Germany
\and
Carnegie Observatories, Las Campanas Observatory, Casilla 601, La Serena, Chile
\and
The Oskar Klein Centre, Department of Astronomy, Stockholm University, AlbaNova, 10691 Stockholm, Sweden
\and
Astronomy Department, University of Illinois at Urbana-Champaign, 1002 W. Green Street, Urbana, IL 61801, USA
\and
Department of Physics and Astronomy, Rutgers, The State University of
New Jersey, Piscataway, NJ 08854, USA
\and
Observatories of the Carnegie Institution for Science, 813 Santa Barbara St., Pasadena, CA 91101, USA
\and
INAF-Osservatorio Astronomico di Padova, vicolo dell Osservatorio 5, 35122 Padova, Italy
\and
George P. and Cynthia Woods Mitchell Institute for Fundamental Physics \& Astronomy, Texas A. \& M. University, Department of Physics, 4242 TAMU, College Station, TX 77843
\and
Center for Mathematical Modelling, Universidad de Chile, Avenida Blanco Encalada 2120 Piso 7, Santiago, Chile
\and
University of Texas at Austin, 1 University Station C1400, Austin, TX 78712-0259, USA
\and
Departamento Ciencias Fosicas, Universidad Andres Bello, Av. Republica 252, Santiago, Chile
}

\date{Received XX XXXX 2014 / Accepted XX XXXX 2014}

\abstract
{
We present ultraviolet through near-infrared (NIR)
  broad-band photometry, and visual-wavelength and NIR spectroscopy of
   Type~Iax supernova (SN)~2012Z.  The data set consists of both
  early and late-time observations, including the first late phase NIR
  spectrum obtained for a spectroscopically classified SN~Iax. 
  A detailed comparison is made to the well-observed Type~Iax SN~2005hk, 
  revealing a striking resemblance between the two objects.
  Simple model calculations of its bolometric light curve suggest
  SN~2012Z produced $\sim$~0.3~$M_{\sun}$ of $^{56}$Ni, ejected 
  about a Chandrasekhar mass 
   of material, and had an explosion energy of
  $\sim$~10$^{51}$ erg, making it one of the brightest ($M_{B} = -18.3$ mag) and most
  energetic SN~Iax yet observed.  Early and late phase
  visual-wavelength spectra of SN~2012Z are found to resemble similar
  epoch spectra of SN~2005hk, with the main difference being the
  former exhibiting broader absorption features, which we attribute to
  variations in the distribution of \nni. The late phase ($+$269d) NIR
  spectrum of SN~2012Z is found to broadly resemble similar epoch
  spectra of normal SNe~Ia; however, like other SNe~Iax, corresponding
  visual-wavelength spectra differ substantially compared to all
  supernova types.  
  Constraints from the distribution of intermediate
  mass elements, e.g. silicon and magnesium, indicate that the outer ejecta did not
  experience significant mixing during or after burning,
  and the late phase NIR line profiles suggests most of the \nni is
  produced during high density burning.   
 The various observational properties of SN~2012Z are found to be 
 consistent with the theoretical expectations of
  a Chandrasekhar mass white dwarf progenitor that experiences a
  pulsational delayed detonation, which produced several tenths of a solar mass of
  \nni during the deflagration burning phase and little (or no) \nni
  during the detonation phase. Within this scenario only a moderate amount 
  of  Rayleigh-Taylor mixing occurs both during the deflagration and fallback phase of the pulsation, 
  and the layered structure of the intermediate mass elements is a product of 
  the subsequent denotation phase.   
   The fact that the SNe~Iax population does not follow a tight brightness-decline relation
  similar to SNe~Ia can then be understood in the framework of
  variable amounts of mixing during pulsational rebound and variable
  amounts of \nni production during the early subsonic phase of expansion. 
}


\maketitle

\section{Introduction}
\label{sec:intro}
Over recent years dedicated non-targeted and targeted transient search
programs have revealed the existence of a multitude of peculiar
transients lurking in the nearby Universe.  A particular group of
objects that share a number of commonalities, warranting their own
spectroscopic designation are Type~Iax supernovae \citep[see][for a review]{foley13}.
Supernova (SN)~2002cx serves as the prototypical example, of what was
originally entitled by \citet{li03} as, ``The Most Peculiar Known
Type~Ia Supernova."  Since the discovery of SN~2002cx, it has become
clear that they are not all that  rare. They have been
estimated to account for  $5 - 30$\% of the overall SNe~Ia rate
\citep{li11,foley13,white14}.  SNe~Iax are characterized by a number of
observational peculiarities, including, among others: a significant
range in peak luminosity ($M_{B}$ $\sim -14$ to $-$18 mag); spectral
signatures characteristic of over luminous 1991T-like SNe~Ia such as
blue continua at early phases and iron lines observed at all epochs;
expansion velocities that are about half (4000 to 9000 km~s$^{-1}$) of what is observed in normal
SNe~Ia (10,000 to 15,000 km~s$^{-1}$) at similar epochs; and late phase optical spectra which bear
little to no resemblance to other types of supernovae.

Detailed studies of the light curves and spectra of well-observed
SNe~Iax  indicate that they generate between $\sim$ 0.003--0.30
M$_{\sun}$ of \nni, and have explosion energies  ranging from $\sim$ 10$^{49}$--10$^{51}$~erg.
The progenitors of the entire class of SNe~Iax are unknown
and are a matter of open 
debate. 
Recent efforts to model  the bright SN~Iax (e.g. SN~2005hk) suggest that 
their origins may lie with the partial thermonuclear incineration of a  carbon-oxygen
Chandrasekhar mass white dwarf (WD), which leaves a bound remnant
\citep{jordan12,kromer13,fink13}. 
However,  pure deflagration models are currently unable to account
for the stratified  distribution of intermediate mass elements (IMEs) that are inferred from the spectroscopic observations of some bright SNe~Iax.
All current deflagration models produce significant mixing throughout
the explosion
\citep{livne93b,livne93a,khokhlov95,g03,R02,g05,roepke07,seitenzahl09,seitenzahl13}, and
it is unclear whether additional physics can sufficiently suppress mixing in 
deflagrations \citep{h13}.
This leaves  open room for other possible progenitor scenarios, such as for example,
the partial thermonuclear disruption of helium accreting
\textit{sub}-Chandrasekhar mass white dwarfs \citep{wang13} or
mergers \citep{Piersanti03}.  
In this paper we discuss an alternative scenario that explains most of the features of the recently observed  Type~Iax SN~2012Z \citep{foley13}, as well as some other members of the  SNe~Iax class.
However, it is important to keep in mind that SNe~Iax are not a homogeneous class and 
several explosion scenarios may contribute to the entire observed population. 

Recently  we studied SNe~2008ha and 2010ae, which are the
least luminous Type~Iax objects yet observed \citep{stritzinger14}.
Here, our attention is turned towards the bright end of the luminosity
distribution of SNe~Iax.  Specifically we present ultraviolet (UV),
optical and near-infrared (NIR) observations of SN~2012Z.  The
data reveal that this object bears a striking resemblance to the
well-observed Type~Iax SN~2005hk
\citep{chornock06,phillips07,sahu08,maund10,mccully14a}, and further indicates
that SN~2012Z is among the brightest and most energetic SN~Iax yet
observed. The majority of the data presented in this study were
obtained as part of the {\it Carnegie Supernova Project-II} (CSP-II),
which is a four year NSF-funded SN followup program aimed at obtaining
optical and NIR observations of $\sim$100 SNe~Ia located in the Hubble
flow.  The CSP-II is placing a particular emphasis on carrying out
detailed NIR spectroscopy, and as a result, this paper presents the
most comprehensive NIR time series of a SN~Iax.  We also note that
revised optical and NIR photometry of SN~2005hk are presented in the
Appendix. Since these data were first published by \citet{phillips07},
the images of SN~2005hk have undergone proper host-galaxy template
subtraction, and as well, the calibration of the local sequence has
been improved.

In a companion paper, we  report on the detection of a 
luminous and blue  
point source coincident with the location  of SN~2012Z in
pre-explosion {\em Hubble Space Telescope} (HST) images
\citep{mccully14b}. 
This direct detection  is consistent with a helium-star donating material to
an accreting white dwarf as the progenitor for SN~2012Z.
Followup HST observations have the potential to determine if the 
progenitor system was correctly identified and consistent with such 
a model. 

The observations of SN~2012Z presented here provide  unique coverage of the
light curve and time evolution of the spectra at 
both optical and NIR wavelengths for a SN~Iax. 
In supernovae, 
we are able to peer into deeper layers of the ejecta 
over time, and therefore, our observations allow us
to probe the entire envelope structure.  
The nature of SNe~Iax is
under discussion, as are many aspects of the explosion physics  and
the nature of the  variations observed within SNe~Ia. A
comprehensive discussion of the latter point is beyond the scope of this
paper and therefore, we refer the reader to 
\citet{h06r}, \citet{h13}, and references therein. 
In what will follow, we discuss the various constraints on the 
underlying progenitor provided by individual observational 
components, as well as compare SN~2012Z to other 
SNe~Ia and SNe~Iax.
As a reference scenario to compare SN~2012Z with
normal SNe~Ia, we adopt one of a series of delayed-detonation models
of the later because it is readily available, fits many observational
properties of SNe~Ia, and can be regarded as a limiting case for one
of the models discussed below.  
The goal of this work is not to rule out a specific model, but rather to piece together the arguments within all scenarios discussed in the literature.
By considering the individual observational constraints of SN~2012Z and confronting them with various  scenarios, we suggest  
a unified picture that may explain 
the nature of at least some SNe~Iax.

\subsection{Supernova 2012Z}

SN~2012Z was discovered with the KAIT telescope as a part of LOSS
\cite[see][]{filippenko01,filippenko05} on 2012 Jan. 29.15 UT, at an
apparent unfiltered magnitude of 17.6 \citep{cenko12}.  With J2000.0
coordinates of $\alpha$ $=$ 03$^{\rm h}$22$^{\rm m}$05\fs35 and
$\delta = -$15$^{\circ}$23$\arcmin$15\farcs6, the position of this supernova
is 14\farcs6 West and 42\farcs1 North from the center of the SA(s)bc
galaxy NGC~1309.  Figure~\ref{FC} contains an image of the host galaxy
with SN~2012Z indicated.  Spectroscopic confirmation obtained three
days after the discovery indicated that this was a SN~Iax  on
the rise \citep{cenko12}.

NED lists the redshift of NGC~1309  as $cz = 2136$~km~s$^{-1}$ or $z=0.007125$.
 NGC~1309 has a number of Cepheid-based distance measurements, and as well has
 hosted the normal Type~Ia SN~2002fk \citep{riess09,cartier14}.
A recent  re-calibration of the distance scale by \citet{riess11}, provides an up-to-date
 Cepheid based distance  to NGC~1309 of  
 33.0$\pm$1.4 Mpc or $\mu = 32.59\pm$0.09. 
Given the accuracy of this distance measurement,  it is adopted  in this paper 
to set the absolute flux scale of SN~2012Z.

\section{Observations}
\label{sec:obs}
\subsection{Ultraviolet, optical, and NIR imaging}

SN~2012Z was observed from space with the UVOT camera aboard the {\it
  Swift} X-ray telescope \citep{burrows05,roming05}.  Up to 13 epochs
of UV $uvw2$, $uvm2$, and $uvw1$-band imaging was obtained over the
duration of a month.  Template images of the host obtained during
early 2013 were used to estimate the background emission at the
position of the SN.  Aperture photometry of the SN was then computed
as described by \citet{brown09}, using the calibration zero-points and
sensitivity corrections from \citet{breeveld11}.  The final UV
photometry is listed in Table~\ref{table:uvot}.
 
Optical ($ugriBV$) imaging of SN~2012Z was initiated near the end of
the first campaign of the CSP-II with the Swope 1.0~m ($+$SITe3 direct
CCD camera) telescope at the Las Campanas Observatory (LCO).  In
addition, NIR ($YJH$) imaging was also obtained with the du Pont 2.5~m
telescope ($+$RetroCam).  All CSP-II imaging was processed in the
standard manner, following the procedures described by \citet{hamuy06}
and \citet{contreras10}.

Before computing photometry of the SN, galaxy subtraction was performed
on all science images. High signal-to-noise optical and NIR 
template images of NGC~1309 were obtained under favorable 
seeing conditions with the du Pont telescope between 
the 10$^{\rm th}$ and the 17$^{\rm th}$ of October 2013. 
Multiple exposures were stacked to create master templates, which were then 
subtracted from the science images following the techniques described by \citet{contreras10}.

PSF Photometry of SN~2012Z was computed differentially with respect to
a local sequence of stars in the field of NGC~1309.  The optical
sequence consists of 17 stars calibrated with respect to
\citet{landolt92} ($BV$) and \citet{smith02} ($ugri$) photometric
standard fields, which were observed over the course of multiple
photometric nights.  The NIR local sequence, on the other hand,
contains 6 stars calibrated relative to \citet{persson98} standards
also observed over multiple photometric nights.  As the
\citet{persson98} catalog of standards does not include $Y$-band
photometry, the $Y$-band local sequence was calibrated relative to
unpublished $Y$-band photometry of a subset of the
\citeauthor{persson98} standards.  The standard star $Y$-band
photometry was computed from multiple observations conducted at LCO as
part of the CSP-I \citep{hamuy06}.  Final absolute photometry of the
optical and NIR local sequence in the {\it natural} system is listed
in Table~\ref{SN12Zlocalsequence}.  The quoted uncertainty for each
magnitude corresponds to the weighted average of the instrumental
errors of the photometry during the nights photometric standard fields
were observed.

With photometry of the local sequences and template subtracted science
images in hand, PSF photometry of SN~2012Z was computed as prescribed
by \citet{contreras10}.  The definitive optical and NIR photometry in
the {\it natural} photometric system of the Swope and du~Pont
telescopes is listed in Tables~\ref{SN12Zoptphot} and
\ref{SN12Znirphot}, respectively.  The uncertainties accompanying each
photometric measurement correspond to the sum in quadrature of the
instrumental error and the nightly zero-point error.  The early phase
UV, optical, and NIR light curves of SN~2012Z are plotted relative to
the time of $B$-band maximum (hereafter $T(B)_{\rm max}$) in
Figure~\ref{earlylcs}.  We note that a single epoch of $BVi$-band
optical photometry was also obtained at late phases. This epoch of
data is used below to provide a lower limit on the UVOIR flux at
$+$266d (see below).

\subsection{Optical and NIR spectroscopy}

Five epochs of early phase ($+$2d to $+$21d), low-resolution
visual-wavelength spectroscopy of SN~2012Z was obtained by the CSP-II
with the Nordic Optical Telescope (NOT$+$Alfosc), and the Magellan
Clay telescope ($+$LDSS3).  In addition, a single high-resolution
optical spectrum was obtained on $-$7d with the Magellan Inamori
Kyocera Echelle (MIKE; \citealt{bernstein03}) spectrograph attached to
the Magellan Clay telescope.  All low-resolution spectra were reduced
in the standard manner using {\tt IRAF}\footnote{The image Reduction
  and Analysis Facility (IRAF) is maintained and distributed by the
  Association of Universities for Research in Astronomy, under the
  cooperative agreement with the National Science Foundation.}
scripts, while the high-resolution spectrum was reduced using the MIKE
pipeline \citep{keelson03} following the methods described by
\citet{simon10}.  Complementing the early spectroscopy are 3 epochs of
late phase ($+$194d to $+$248d), low-resolution spectra, which were
obtained with SALT ($+$ RSS), the du~Pont ($+$WFCCD) and Keck
($+$DEIMOS) telescopes.

A summary of our spectroscopic observations and those of
\citet{foley13} is provided in the spectroscopic journal given in
Table~\ref{specjor}.  As seen in Figure~\ref{optspec}, our early phase
spectra nicely complement the previously published spectra of
SN~2012Z.  The three late phase visual-wavelength spectra are plotted
in Figure~\ref{nebopt}.

Ten epochs of NIR spectroscopy were also obtained for SN~2012Z, and
details concerning these observations are provided in
Table~\ref{specjor}.  This includes 5 epochs taken with the Magellan Baade
telescope ($+$FIRE), 4 epochs with the Very Large Telescope
(VLT$+$ISAAC), and finally, one epoch with the New Technology
Telescope (NTT$+$Sofi). The FIRE spectra were reduced as described by
\citet{hsiao13}, while the VLT and NTT spectra were reduced as
described by \citet{smartt13}.  The NIR spectroscopic sequence plotted
in Figure~\ref{nirspec} extends from $-$7d to $+$269d, and consists of
the most comprehensive NIR time-series yet obtained for a SN~Iax,
including the first late phase NIR spectrum.

 \section{Dust extinction}
\label{hostproperties}

The Milky Way color excess in the direction of NGC~1309 is listed in
NED to be $E(B-V)_{\rm MW} = 0.035\pm0.012$ mag \citep{schlafly11},
which when adopting a \citet{fitzpatrick99} reddening law
characterized by an $R_{V} = 3.1$, corresponds to $A_{V} = 0.11$ mag.

Estimating the level of reddening that affects the emission of supernovae due
to dust external to the Milky Way (MW) is a challenging prospect,
particularly for peculiar types that do not have well-defined
intrinsic colors.  One must therefore find alternative methods to
estimate host reddening.  In the case of SN~2012Z, we first examine
the host-galaxy color-excess value inferred from the broad-band
optical and NIR light curves of the normal Type~Ia SN~2002fk, which
occurred in an inner region of NGC~1309.  From a detailed analysis of
its broad-band colors, \citet{cartier14} estimate the color excess
along the line-of-sight of SN~2002fk to be $E(B-V)_{host} =
0.057\pm0.036$~mag.  As SN~2012Z is located in the outskirts of
NGC~1309 one may expect that it suffered a similar level or even less
reddening.

Using the high-resolution MIKE spectrum, we next explored if a more
direct handle on the host reddening could be obtained via column
density measurements of neutral sodium and potassium inferred from
host absorption components of the $\ion{Na}{i}~D$
$\lambda\lambda$5890, 5896 and $\ion{K}{i}$ $\lambda\lambda$7665, 7699
doublets.  Line fits computed using the Voigt profile fitting program
\href{http://www.ast.cam.ac.uk/~rfc/vpfit.html}{\tt
  VPFIT}
provided column densities of log$_{10}$ $N_{\ion{Na}{i}} =
12.79\pm0.09$~cm$^{-2}$ and log$_{10}$ $N_{\ion{K}{i}}=
11.30\pm0.09$~cm$^{-2}$.  These values give a $N_{\ion{Na}{i}}$ to
$N_{\ion{K}{i}}$ ratio that is perfectly normal for the MW
\citep[see][Figure~13]{phillips13}.  Adopting the MW relations between
log$_{10}$ $N_{\ion{Na}{i}}$ and log$_{10}$ $N_{\ion{K}{i}}$
vs. $A_{V}$, as derived by \citet[][see their Equations~4 and
5]{phillips13}, implies visual extinction values of
$A_{V}(\ion{Na}{i}) = 0.45\pm0.24$ mag and $A_{V}(\ion{K}{i}) =
0.50\pm0.36$ mag, respectively.

\citet{phillips13} also recently showed that the 5780~\AA\ diffuse
interstellar band (DIB) can provide a robust means to estimate host
reddening. Under close inspection of the high-resolution spectrum of
SN~2012Z this feature is not detectable. However from the
signal-to-noise ratio of the MIKE spectrum at the expected position of
this feature, a 3$\sigma$ upper limit on the equivalent width of
EW(5780) $\leq$ 35 m\AA\ is estimated.  Adopting the \citet[][see
their Equation~6]{phillips13} MW relation that connects log$_{10} EW
(5780)$ to $A_{V}$ then implies an $A_{V} \leq 0.18\pm0.09$~mag.  This
value and those derived from $\ion{Na}{i}$ and $\ion{K}{i}$ doublets
are generally consistent, and when combined via a weighted average
imply an $A_{V} = 0.23\pm0.08$ mag, which when adopting an $R_V =
3.1$, corresponds to a host galaxy color excess $E(B-V)_{host} =
0.07\pm0.03$~mag.  Interestingly, this value is fully consistent with what is
derived from the broad-band colors of SN~2002fk.  In what follows we
adopt this value for the host reddening of SN~2012Z, and when
combining with the MW component, gives the total color excess
$E(B-V)_{tot} = 0.11\pm0.03$~mag.

\section{Results}
\label{sec:results}

\subsection{UV, optical and NIR light curves}
\label{sec:lcs}

The photospheric phase optical light curves of SN~2012Z plotted in
Figure~\ref{earlylcs} consist of dense photometric coverage, while the
UV light curves are of considerably lower quality, and the NIR light
curves contain only a handful of epochs, beginning well after maximum.
The optical light curves reveal the characteristic bell-shape
appearance that is typical of stripped-envelope supernovae, indicative of a
rather low ejecta mass.  Similar to other bright SNe~Iax, the $r$- and
$i$-band light curves appear broad and slowly evolving as compared to
the bluer bands.  In addition, the $r$- and $i$-band light curves show
no evidence for a secondary maximum, which is a defining
characteristic of normal SNe~Ia \citep{hamuy96}.  Interestingly, the
$uvw2$- and $uvm2$-band light curve are found to be already declining
8 days before $T(B_{max})$, while the $uvw1$ light curve appear to be
at peak brightness. Clearly, the UV light curves reach maximum well
before the optical light curves.
 
Light curve parameters were measured from Gaussian Process functional
fits to the optical light curves using software contained within the
light-curve fitter SNooPy \citep{burns11}.  The fit parameters consist
of: (i.) the time of maximum, (ii.) peak apparent magnitude and,
(iii.) an estimate of the decline-rate parameter,
$\Delta$m$_{15}$\footnote{The light curve decline-rate parameter,
  $\Delta$m$_{15}$, is defined as the magnitude change from the time
  of maximum brightness to 15 days later. For normal SNe~Ia,
  $\Delta$m$_{15}$ correlates with the peak absolute luminosity in
  such a way that more luminous objects exhibit smaller
  $\Delta$m$_{15}$ values \citep{phillips93}, however this is not the
  case for SNe~Iax (see e.g., Figure~\ref{dm15vslum}).}.  These fit
parameters are listed in Table~\ref{lcpar}, along with those measured
from the revised photometry of SN~2005hk (see Appendix), accompanied
with robust uncertainties derived via Monte Carlo simulations. 
The
light-curve fit parameters indicate that the bluer the light curve the
earlier maximum is reached, with a $\Delta$t $\approx$ 13 day
difference between the time of $u$- and $i$- band maxima, and the
redder the light curve the smaller its decline-rate.  The decline-rate
ranges from as much as 1.92$\pm$0.08 mag in the $u$ band, all the way
down to 0.54$\pm$0.04 mag in the $i$ band.

Figure~\ref{optcolors} shows the $(B-V)$, $(V-r)$ and $(V-i)$
color evolution of SN~2012Z compared to that of the Type~Iax SN~2005hk
and the normal, un-reddened Type~Ia SN~2006ax.  The color curves of
both SNe~Iax have been corrected for MW and host reddening.  Overall
the colors and the color evolution of SNe~2005hk and 2012Z are very
similar, with the former appearing marginally bluer.  At the start of
the observations the color curves of SN~2012Z are at their
bluest. They then slowly evolve and reach a maximum value in the red
around 20 days past $T(B)_{max}$, at which point an inflection is
observed followed by a linear evolution with a negative slope.  The
color evolution of the two SNe~Iax also have a similar overall shape to
normal SNe~Ia, but with colors that are consistently redder as the
supernovae evolve.  It will be interesting to see in the future, as more
well-observed, bright SNe~Iax become available, if this family of
objects displays intrinsic colors or color evolution which could allow
for an accurate estimation of the host extinction.
  
\subsection{Absolute magnitudes, UVOIR light curve, and light curve modeling}
\label{sectionuvoir}

With estimates of peak apparent magnitude for the $ugriBV$ bands in
hand, peak absolute magnitudes were computed adopting an $E(B-V)_{tot}
= 0.11 \pm 0.03$~mag and the direct distance measurement $\mu =
32.59\pm0.09$~mag.  The resulting peak values are listed in
Table~\ref{lcpar}\footnote{To derive absolute magnitudes of SN~2005hk
  (and absolute luminosity, see Section~\ref{sectionuvoir}), we have
  adopted a distance modulus of $\mu = 33.46\pm0.27$ mag
  \citep{phillips07} and the total color excess value $E(B-V)_{host} =
  0.11$ mag.}, with accompanying uncertainties that account for both
the error in the fit to the time of maximum, and the error associated
with the distance to the host galaxy.  Reaching an absolute peak
$B$-band magnitude ($M_B$) $\approx$ $-18.3$$\pm$0.1 with
$\Delta$$m_{15}(B) = 1.43\pm0.02$ mag, SN~2012Z is among the brightest
and slowest declining SN~Iax yet observed.  
This is demonstrated in
Figure~\ref{dm15vslum}, where the location of SN~2012Z in the $M_B$
vs. $\Delta$$m_{15}$ diagram is shown compared to a handful of other
well-observed SNe~Iax, and an extended sample of SNe~Ia observed by
the CSP.
  
Armed with the broad-band photometry of SNe~2005hk and 2012Z, we
proceeded to construct UVOIR bolometric light curves.  For SN~2005hk
we adopted the early phase photometry presented in Appendix~A, as well
as the late phase photometry published by \citet{sahu08}.  The early
phase filtered light curves of both objects were fit with spline
functions, which allowed us to fill in missing gaps in the data based
on interpolation.  For SN~2012Z the $u$-band light curve was extended
so that its temporal coverage matched that of the other optical
filters. To do so, we resorted to extrapolation by adopting an average
($u-B)$ color derived from the last three epochs in which $u$-band
photometry was obtained.

With the definitive observed light curves in hand, the photometry was
corrected for extinction, and then converted to flux at the effective
wavelength of each passband. This allowed us to construct SEDs
spanning from $\sim$~300 to 2500~nm.  The full series of SEDs were
then summed over wavelength using a trapezoidal integration technique,
assuming zero flux beyond the limits of integration.  Due to our
limited NIR photometric coverage of SN~2012Z it was necessary to
account for the NIR contribution of the flux.  Given the close
resemblance between the peak luminosity, light curve shape (see Section~\ref{sec:compobj}), and color
evolution between SNe~2005hk and SN~2012Z, we resorted to adopting the
fraction of NIR flux estimated from the full optical/NIR spectral
energy distribution (SED) of SN~2005hk.  The final UVOIR light curves
of SNe~2005hk and SN~2012Z are plotted in Figure~\ref{uvoir}, with the
associated error-bars accounting for uncertainties in the photometry,
total reddening, and distance.  Note that also over plotted in
Figure~\ref{uvoir} are a handful of flux points of SN~2012Z, which
were computed using its optical {\em and} limited NIR photometry.
These points are in full agreement with the UVOIR flux points of
SN~2012Z that contain the contribution of NIR flux estimated from
SN~2005hk.

To derive rough estimates of the explosion parameters for these
SNe~Iax, we turned to toy model fits of the UVOIR light curves based on
Arnett's equations \citep{arnett82}, following the implementation
described by \citet{valenti08}.  The key parameters of interest are
the amount of radioactive $^{56}$Ni synthesized during the explosion,
the ejecta mass ($M_{ej}$), and the explosion energy ($E_K$).  The
model calculations rely upon a number of underlying assumptions and
therefore the results should be approached with caution.  With this
caveat in mind, the model assumptions include: 
complete energy deposition supplied from the radioactive decay chain
$^{56}$Ni$\rightarrow$$^{56}$Co$\rightarrow$$^{56}$Fe,
spherical symmetry, homologous expansion of the ejecta, no appreciable mixing of
$^{56}$Ni, a constant optical opacity (i.e. $\kappa$$_{opt} ~=~$0.1
cm$^{2}$~g$^{-1}$), and the diffusion approximation for photons.  A
key input parameter in computing the model calculations is a measure
of the expansion velocity ($v_{ph}$) of the ejecta. 
From a detailed inspection of the optical spectral features of SN~2012Z  (see Section~\ref{sec:earlyspectra}), this parameter is estimated to range between
$7,000 \lesssim v_{ph} \lesssim 9,000$ km~s$^{-1}$.  In the case of
SN~2005hk, based on velocity measurements presented by
\citet{phillips07} and \citet{mccully14a}, we infer the range $5,000
\lesssim v_{ph} \lesssim 7,000$~km~s$^{-1}$.  For both objects a grid of
models was computed that encompasses the appropriate range of
$v_{ph}$.  Final best-fit model calculations are over-plotted in
Figure~\ref{uvoir}.  The range of explosion parameters computed for
SN~2005hk are $M_{Ni} = 0.15-0.20~M_{\sun}$, $M_{ej} =
1.5-2.0~M_{\sun}$ and $E_{k} = 0.4-1.2\times10^{51}$~erg; while those
of SN~2012Z are $M_{Ni} = 0.25-0.29~M_{\sun}$, $M_{ej} =
1.4-2.6~M_{\sun}$ and $E_{k} = 0.7-2.8\times10^{51}$~erg.

These rough estimates of the explosion parameters are in essence 
based on Arnett's Rule and the the photospheric expansion velocity, \vphot. 
It should be stressed that a portion of the large range in 
the estimated parameters is due to the degeneracy between the $E_{k}$, $M_{Ni}$ and $M_{ej}$, and the reliance on, in effect, a single bolometric light curve. 
Furthermore, we note that the estimated values of the various explosion parameters 
do not capture enough of the physical variations displayed by the various suite of SNe~Ia explosion
models found in the literature.
Even for normal SNe~Ia, Arnett's Rule is but a useful means of
  obtaining a qualitative understanding of the relation between the
  mass of \nni produced and the peak brightness. 
 In particular, due to the strong temperature dependence of the
 opacity, opacities vary by two and a half orders of magnitude {at each phase of  the evolution of the supernova} \citep{hkm92}. 
 This opacity
 variation is required in order to reproduce the brightness decline-rate
 relation \citep{HKWPSH96,nugseg95,PE01,maeda03,baron12}.
    Attempts to account for both gamma-rays escape and energy stored within the 
  expanding ejecta due to variations in the diffusion time and
  non-constant opacities,  have led to further  parameterization to the
  standard form of Arnett's Rule.  
  This is normally accomplished by the introduction 
  of the parameter, $\alpha$, which leads to
  $L_{\text{bol}}^{\text{max}} \approx \alpha \dot S(t_{\text{max}})$. 
  The parameter $\alpha$ is thought to  lie somewhere in the range $0.7 \la \alpha \la 1.5$ 
  \citep[e.g.][]{branch92,khokhlov93,hk96}.\footnote{In the reference model adopted in this paper $\alpha = 1.22$.}
  With such a large range in $\alpha$, attempts to use Arnett's Rule
for quantitative analysis will only be accurate to about 50\%.

SNe~Iax light curves are characterized by a short rise time
  as compared with normal SNe~Ia.
The hydrodynamics is determined by the total ejected mass and
the total energy produced by thermonuclear fusion; however, the light curves
are determined not by the total mass, but rather by the optical
depths, that is the product of the column density and the
opacities. As the opacities vary by several orders of magnitude over
time as well as varying with abundances and ionization stage,
a change of the opacity by a factor of two is
equivalent to a variation in the mass by $\sim\sqrt{2}$. 
Choosing a
constant opacity to fit with the risetime is equivalent to fixing the
mass scale and the opacity choice of Arnett's Rule is tuned to the
choice of a Chandrasekhar mass.

In Section~\ref{sec:analysis}  we attempt to gain an additional handle on
the explosion parameters of SN~2012Z through comparison of the observations to modern
explosion models.

\subsection{Early phase spectroscopy}
\label{sec:earlyspectra}

The sequence of early phase visual-wavelength spectra plotted in
Figure~\ref{optspec} reveals all of the quintessential features of a
SN~Iax located at the bright end of the luminosity distribution.  At
the earliest epochs, SN~2012Z exhibits a ``hot" continuum dotted with
prevalent iron features.  These characteristics are reminiscent of
early spectra of luminous 1991T-like SNe~Ia, however, they
differ substantially at late phases. 
The blue portion of
the $-$9d spectrum exhibits clear features produced by $\ion{Ca}{ii}$
$H\&K$, $\ion{Fe}{ii}$ $\lambda$4555 and $\ion{Fe}{iii}$
$\lambda\lambda4404, 5129$, while red-wards of $\sim$ 5000~\AA\ the
spectrum is relatively smooth, showing only subtle notches at the
expected locations of the IMEs
$\ion{S}{ii}$ $\lambda\lambda5454, 5640$ and $\ion{Si}{ii}$
$\lambda6355$.  By maximum, the strengths of these IME features mildly
increase, however soon thereafter, their presence is diluted by the
emergence of numerous permitted lines of $\ion{Fe}{ii}$ and
$\ion{Co}{ii}$ \citep[see][for detailed SYNOW fits to similar epoch
spectra of SN~2002cx]{branch04}.  Simultaneously, a conspicuous
absorption feature attributed to the $\ion{Ca}{ii}$ NIR triplet
emerges, and subsequently grows in strength over the duration of a
fortnight.  Finally, we note that the near maximum spectra of SN~2012Z
shows a  weak feature that has been attributed to $\ion{C}{ii}~\lambda6580$ 
\citep[see][for a 
discussion]{foley13}.

The NIR spectral time-series of SN~2012Z presented in
Figure~\ref{nirspec} contains both the earliest and latest observed
epochs of any SN~Iax published to date, and therefore provides unique
insight into the SN~Iax class.  Similar to normal SNe~Ia observed at very 
early times, the
earliest epoch ($-7$d)  NIR spectrum of SN~2012Z exhibits a rather blue,
featureless continuum.  Several days later, just prior to maximum
 the spectrum reveals a shallow feature at
wavelengths expected for
$\ion{Mg}{ii}~\lambda1.0952$ $\mu$m. 
There is also weak evidence of 
$\ion{C}{i}$ lines at $0.9093$,
$0.9406$, $1.0693\ \mu$m, which have been clearly identified in the 
subluminous Type~Ia SN~1999by \citep{hoeflich02a}.
As the SN ejecta
expands, the photosphere recedes into the inner regions, resulting in
a reduction of the continuum flux, while line features largely associated
with Fe-group elements emerge.  The evolution of the spectrum and the
emergence of Fe-group features tends to occur on a faster timescale in
SNe~Iax as compared to normal SNe~Ia.

Close inspection of the post maximum spectra reveals that by $+$22d,
$\ion{Co}{ii}$ features dominate the NIR spectrum red-wards of
1.5~$\mu$m. This is expected, owing to the fact that $\ion{Co}{ii}$
features are known to be ubiquitous to the NIR spectra of all other
SNe~Iax observed to date \citep[see][]{stritzinger14}.  Turning to the
last two observed NIR spectra, minimal evolution of the overall line
features is observed to occur between $+$58d and $+$269d.  Finally, we
note that the NIR spectra of SN~2012Z do not exhibit any compelling
evidence for spectral features related to $\ion{He}{i}$ \citep{white14}.

The time-evolution of the blue-shifts of a handful of ions present in
the early spectra of SN~2012Z were measured in order to estimate the
velocity distribution of the emitting regions of the ejecta.  The
results of which are shown in Figure~\ref{velocities}, and include IME
ions of $\ion{Ca}{ii}$~$H\&K$, $\ion{Si}{ii}$~$\lambda6355$, 
 $\ion{C}{ii}$~$\lambda6580$ and
$\ion{Ca}{ii}$~$\lambda8498$, as well as the iron lines $\ion{Fe}{ii}$
$\lambda$4555 and $\ion{Fe}{iii}$~$\lambda4404$, $\lambda5129$.  The
expansion velocity measurements reveal that these ions are related to material with relatively
low velocities, all consistently below 10,000~km~s$^{-1}$.
Interestingly, at $+$2d, ions attributed to $\ion{Ca}{ii}~H\&K$,
$\ion{Fe}{iii}~\lambda4404$ and $\ion{Si}{ii}~\lambda6355$ exhibit
consistent line velocities of $\approx$ 7,500~km~s$^{-1}$. This value
is bracketed by velocities of $\ion{Fe}{ii}~\lambda4555$ and
$\ion{Fe}{iii}~\lambda5129$, which are $\sim$ 1,000~km~s$^{-1}$ higher
and lower, respectively.  Inspection of Figure~\ref{velocities} also
indicates that at a month past $T(B)_{max}$ the line velocities evolve
rather slowly, with velocities ranging between 6,000 and
8,000~km~s$^{-1}$. 
Taken together the line velocities inferred from the
spectroscopic sequence of SN~2012Z suggests a rather low explosion
energy compared to normal and  over-luminous SNe~Ia.

To gain a handle on the radial structure of IMEs in the outer
  layers of the ejecta, 
 we measured the blue wing velocities in addition to the absorption minima. Blue wings are important since
 they will not appear if the particular ion does not exist at that velocity.
 In particular we measured the velocities  of the $\ion{Si}{ii}$ $\lambda6355$ and $\ion{Mg}{ii}$ 
$\lambda{1.0952}$~$\mu$m features. Metal lines are not appropriate
since even primordial abundances of calcium and beyond can produce
background lines. 
 The  $\ion{Si}{ii}$ blue wing is measured to be $v \approx -11,000$~\kmps  on  $-$8d and 
 $v \approx -10,000$~\kmps on $+$1.8d, while $\ion{Mg}{ii}$ is measured with $v \approx -13,500$~\kmps  on $-0.4$d. 
  These measurements along with the flat evolution of features in velocity space  suggests a layered distribution of the IMEs in the outer ejecta 
 \citep{nomoto84}. Similar abundance structures have been 
 inferred from the spectral analyses of normal SNe~Ia 
  \citep{JB90,hkm95,hk96,baron06,dessart14b},
 and provides  robust constraints on the explosion physics (see below).

 \subsection{Late phase spectroscopy}
\label{lateir}
The late-phase visual-wavelength spectra of SN~2012Z (see Figure~\ref{nebopt}) contains
nearly a dozen broad emission features, with the majority of cooling
occurring through the most prevalent feature centered around
$\approx$~7400~\AA.  This dominant emission feature peaks by a factor
of 3 higher in flux than any of the other emission feature, and has a
full-width-half-maximum (FWHM) velocity $\sim 10,000$~\kmps.  For
comparison the weaker emission features exhibit FWHM values between
5,700 to 9,000~km~s$^{-1}$.  Based on similar epoch spectra of other
SNe~Iax that have lower ejecta velocities, we attribute the
$\approx$~7400~\AA\ feature to a blend of a handful of lines,
including $[\ion{Ca}{ii}]$~$\lambda$$\lambda$7291, 7324 and
$[\ion{Fe}{ii}[$~$\lambda$7155.  \citet{foley13} remarked that
\textit{some} SNe~Iax exhibit $[\ion{Ni}{ii}]$ $\lambda7378$ and 
[$\ion{Fe}{ii}$] $\lambda$7453; unfortunately in SN~2012Z the
significant blending 
of features in this wavelength range prevents simple line
identifications and therefore, it not straightforward to determine if
these lines are present.  Other notable features in the late phase
spectra located at $\sim$~5200~\AA\ and 5400~\AA\ are likely produced
by the blending of numerous permitted $\ion{Fe}{ii}$ lines, which have
been identified in other SN~Iax \citep{jha06,foley13}.  
{The emission feature at $\approx$~5900~\AA\ is likely formed from 
$\ion{Na}{i}~D$, while at $\approx$~8300~\AA\ we find features 
which are  more than likely formed by the $\ion{Ca}{ii}$ NIR triplet. 
The NIR triplet is commonly seen in SNe~Ia and other supernovae, 
and is strong even at solar abundances. 
An alternative identification of the $5900$ \AA\ feature could be 
$[\ion{Co}{iii}]$ $\lambda$$5888$ \citep{dessart14a}. 
Note that we find a blend of $[\ion{Co}{iii}]$  $\lambda5890$, and contributions of
$[\ion{Co}{iii}]$ $\lambda5908$ and $[\ion{Fe}{ii}]$  $\lambda5872$ in our reference model, but much weaker than observed (see Section~\ref{nebspectra}).

Beyond these  prevalent features, the late phase spectrum of SN~2012Z 
contain a multitude of low amplitude and low velocity (500-1000 km~s$^{-1}$) features.
These features appear to be generic  to the SNe~Iax class, and
have been linked to  both forbidden and allowed transitions of iron
\citep[e.g.][]{jha06,mccully14a}.

We now turn to the examination of the late phase NIR spectrum of SN~2012Z, which to date, is the first of its kind obtained for a SN~Iax.
Plotted in Figure~\ref{LTnirscomp} is the comparison between our 
 late phase  ($+$269d) NIR spectrum of SN~2012Z, and
 a similar epoch spectrum of the normal Type~Ia SN~2012fr.
 Clearly, SN~2012Z exhibits many of the same broad features as is seen in the normal 
 SN~Ia, particularly at the wavelength regions that correspond to the $J$ and
$H$ passbands.
Overall, the similarity between the two spectra is quite striking considering the 
 discrepancy between the visual-wavelength late phase spectra between 
 SN~2012Z and other SNe~Iax in general, compared to normal SNe~Ia
 (see Figure~\ref{nebcomp}).
Close inspection of the late phase NIR spectrum, albeit of low
  signal-to-noise,  indicates that the 
$[\ion{Fe}{ii}]$ $\lambda$1.64~$\mu$m feature has a pot-bellied profile rather than
a peaked profile.\footnote{There is confusion in community in the nomenclature
  to describe central line profiles. Line 
  profiles from constant emissivity
  shells which are truly flat-topped have been called by the same
  name as those due to a nickel hole where energy input from gamma
  rays and positrons causes the emissivity to vary. 
  While flat-topped has previously been used in the literature, we want to
  avoid the confusion between the description of truly
  flat-topped shell emission that is used to describe optically
  thin line profiles in a moving shell \citep{OC01} and the profiles
  we are discussing here, which we refer to as pot-bellied.}
Similar features have been previously observed in a handful of 
normal SNe~Ia \citep[e.g.][]{hoeflich04,motohara06}, and as we discuss below, is a signature of high density burning.

\section{Bridging observations to model calculations}
\label{sec:analysis}

In the subsequent section, we use the various observational properties of SN~2012Z 
described in the previous section as a guide to argue that the progenitor is consistent with the 
theoretical expectations of a pulsating delayed detonation (PPD) explosion of a 
Chandrasekhar mass ($M_{\text{Ch}}$)  WD.  
To do so, we rely upon existing explosion models and   
 link together model predictions to the various observables  including amongst others: 
 (1) the light curve properties, (2) the layered structure of the ejecta and, (3) the pot-bellied $[\ion{Fe}{ii}]$ 1.64~$\mu$m  profile. 
The various lines of reasoning are then connected and summarized in 
 Section~\ref{sec:discussion} to provide a progenitor scenario to explain
 at least some SNe~Iax.

\subsection{Interpreting the broad-band emission through model comparison}

SNe~Ia explosions produce a form of stellar amnesia \citep{h06r} and
consequently, most explosion scenarios produce very similar structures,
ranging from classical deflagration models like W7 to the inner
structures of merger  \citep{nomoto84,benz90,livio89,khokhlov91b,yamaoka92,hk96,loren09,g03,g05,roepke07,seitenzahl09,pakmor10}.
The abundance distributions reflect the conditions under which burning
takes place and therefore exhibit the imprint of the detailed physics that
we discuss below.  
Delayed detonation models (DDT) give the overall chemical stratification  where the  burning propagates as  a deflagration front and then  transitions into a 
detonation \citep{khokhlov91b,yamaoka92}.
Within the DDT scenarios there are two realizations: (1)  the WD becomes unbound during the deflagration phase (these are the DD models) and, (2) 
the WD remains bound at the end of the deflagration phase, then undergoes a 
pulsation followed by a  delayed detonation transition. 
This latter scenario is called a pulsational delayed detonation (PDD). 
These two realizations form a sequence which merge at the regime where the WD just becomes unbound.
In what follows  a classical DD model is  used as reference, as well as a PDD model with a large amplitude pulsation to span the range \citep{hkm95,hk96}.
Beyond our own past work on PPD models, we note that
 \citet{dessart14b} have recently  presented a suite of PDD models that cover the range of observed SNe~Ia luminosities. Their results are similar to those of  
 \citet{hkm95,hk96}, and show that PDDs can encompass a wide variety
 of SNe~Ia conditions. 
 
 We identified a set of simulations based on the  spherical explosion of a 
$M_{\text{Ch}}$  WD, which inherently suppresses mixing.
One particular simulation within a series of  DD models  was identified to exhibit the approximately correct brightness of SN~2012Z.
 This particular simulation, designated 5p0z22.16 \citep{hoeflich02a} is a DD
model in which a subsonic deflagration burning front turns into a supersonic detonation wave and mixing in the deflagration phase is suppressed. 
The 5p0z22.16 simulation gives $[M_B,(B-V)_{max},\Delta m_{15}(B),\Delta m_{15}(V)]$
of $[-18.22, 0.14, 1.54, 0.99]$, which compares well to $[-18.00, 
  0.07, 1.62, 0.92]$ and $[-18.27, 0.01, 1.43, 0.89]$
for SN~2005hk and SN~2012Z, respectively. 
In addition, the model has total and \nni masses of
$1.362$ and $0.268$~\msol, respectively, an explosion energy of
$1.21\times 10^{51}$~ergs, and also produces light curves where the primary and secondary NIR maxima merge into a single maximum \citep{hkm95,hoeflich02a}.

From the  observed light curves, $M_{\text{Ch}}$  models are consistent with SN~2012Z
within the uncertainties of $\approx 10$\% for different
progenitors and $\approx 20$\% for different explosion scenarios \citep[][and
references therein]{hk96,h13}.  
 However, in 5p0z22.16 the photosphere recedes from
a velocity of about $13,000$~\kmps on $-10$d to about $9,000 $~\kmps
by maximum light, which is too high compared to what is observed 
in SN~2012Z.  
Light curves capture the diffusion time scales of the energy production, but, as
discussed below, it takes line profiles to probe the velocity
distribution of the elements to understand the nature of SNe~Iax.

\subsection{Constraints from early phase spectroscopy}

In what follows   5p0z22.16 is used as a reference model,
but we must consider variations from it that can explain the features of SNe~Iax in general, and SN~2012Z in particular.  Namely, the model predictions of the photospheric velocities must match the observations.
 Although this model is a classical DD model, we can think of it as the first  in
a series of PDD models \citep{hkm95}, but with a shell mass of zero. 
Below we consider the effects of variations on the shell
mass and how this allows us to subdue the photospheric velocities to match 
those observed in SN~2012Z.
 We will be guided in our considerations by the results shown in
Figure~10 of \citet{quimby07}.
Due to the low density in the shell some of the original WD's  carbon and oxygen (C$+$O) 
remains completely unburned in the outer layers. This is in distinction to the classical DD model where all of the material undergoes at least carbon burning. 

Early-time spectra can be used to probe the decoupling region of the photons
and give a unique measure of the velocity distribution of the elements.
Thus, spectral observations probe the density under which burning took place.  
In velocity space, the 5p0z22.16 simulation \citep[see Figure 3 in][]{hoeflich02a} shows products of explosive carbon burning, namely oxygen and magnesium, incomplete oxygen burning, that is silicon and sulphur,
and complete burning close to nuclear statistical equilibrium (NSE), i.e. 
iron/cobalt/nickel (Fe/Co/Ni), in the velocity ranges of
$[13,000-18,000]$, $[7,000-14,000]$ and $[0-6,000]$~\kmps,
respectively. An inner region of about $3,000$~\kmps shows elements
of stable isotopes of Fe/Co/Ni due to electron capture, a signature of
high-density, $M_{\text{Ch}}$  models. Calcium is produced in the range
$[5,000-13,000]$~\kmps, and is destroyed at lower velocities.

 In the standard  DD scenario for subluminous SNe~Ia, specifically
 model 5p0z22.16, deflagration burning with mixing during the deflagration suppressed, only burns material up to velocities of about 
 $4,300$~\kmps, \citep[see e.g., Fig.~1 in][]{hoeflich02a}.
SN~2012Z shows Fe-group elements up to, at least, $9,000$~\kmps
(Figure~\ref{velocities}), which is some $3,000$~\kmps higher than where
\nni is produced during the detonation phase of the reference model. 
Fe-group elements are formed during high density burning beyond 
1-2$\times10^{7}$~\gcm, strongly suggesting mixing in a $M_{\text{Ch}}$ scenarios, or an increase in production of \nni during the detonation phase.

An increased production of \nni within the classical-detonation
scenario would lead to an even more extended region of IMEs 
 than 5p0z22.16, which is inconsistent with SN~2012Z. Detonations in smooth density structures will unavoidably produce extended distributions of IMEs in velocity space or if the density is high,  hardly any production of IMEs.
Because the specific energy production depends on the nuclear processes, the latter case would also produce high velocity \nni \citep{khokhlov93,hk96}.  
We note that the same arguments would apply to sub-$M_{\text{Ch}}$ explosions triggered by the ignition of an outer helium layer
\citep{wwt80,nomoto82a,ww94,hk96,nughydro97,sim10}.

Mixing may occur both during the deflagration phase of burning and during a pulsation phase of the WD prior to the initiation of the detonation, while the detonation phase leads to a layered structure.
During the last decade, significant progress has been made in our
understanding of the deflagration phase.  
On a microscopic scale, a deflagration propagates due to heat conduction by electrons. 
However, all current simulations have found burning fronts to be Rayleigh-Taylor (RT) unstable increasing the effective burning speed, predicting homogeneous mixing of products of incomplete and complete burning \citep{livne93b,livne93a,khokhlov95,g03,R02,g05,roepke07,seitenzahl09,seitenzahl13}.  
As seen in Figure~\ref{velocities}, mixing of IMEs and calcium is not observed in SN~2012Z at
velocities less than $\approx 6,500$~\kmps. 
Ca~II is commonly seen in SNe~Ia even in the outer layers, that at most undergo carbon burning
at solar metallicities \citep[see for example][]{lentz00}, arguing against mixing of the inner layers during a deflagration burning phase.
 
Both mixing and an upper limit for the distribution of burning products can be understood in terms of models that undergo a pulsational phase prior to the detonation \citep{khokhlov93,hoeflich04,hk96,livne99,livne05,quimby07,dessart14b}. 
In this class of PDD models, a $M_{\text{Ch}}$ WD remains bound after the deflagration
phase. 
The deflagration mostly produces Fe-peak isotopes; subsequently, an extended envelope is formed.  
As the explosion progresses, and the bound shell falls back a detonation is triggered that
leads to the production of IMEs and a massive shell. 
The mass of the shell determines the upper velocity limit of the burning products \citep{khokhlov93,hk96,quimby07}.  
To be aligned with SN~2012Z's measured blue wing velocity 
of $\ion{Si}{ii}$ (see Section~\ref{sec:earlyspectra}),
a shell mass of $\approx 0.05-0.12$~M$_\odot$ is 
required \citep[see Figure~10 in][]{quimby07}.  
In addition the blue wing velocity of $\ion{Mg}{ii}$ is consistent with a shell mass of $\sim 0.08$~\msol.
Early multi-dimensional calculations indicated that WDs become unbound during the deflagration
phase \citep{g03,roepke07}, but this conclusion depends sensitively on the assumption for the thermonuclear runaway \citep{livne05}.  
This strong dependency on the basically unknown initial conditions of the WD is an obstacle for fitting details of a specific supernova, but is consistent for the wide variety of properties displayed by SNe~Iax.

It is well known that weak interaction/pulsation or the presence of
a low-mass buffer around the exploding white dwarf can dramatically
alter the outer ejecta layers of the SN Ia explosion  
\citep{hkm95,hk96,dessart14b}.  
The specific properties of the pulsation can affect
the appearance of $\ion{C}{i}$, $\ion{C}{ii}$, and $\ion{O}{ii}$ lines 
The presence or absence of  $\ion{C}{i}$, $\ion{C}{ii}$ and $\ion{O}{ii}$ 
is  also sensitive to the details of the excitation. 
Some realizations within PDD and DD models  do in fact show strong 
optical and NIR $\ion{C}{i}$ and $\ion{C}{ii}$ lines \citep{hoeflich02a,dessart14b}. However, $\ion{C}{ii}$ $\lambda$6580 lines at velocities as
$\sim 7,000$ \kmps  (Fig.~\ref{velocities}) is somewhat in
contradiction with also weak $\ion{C}{i}$ $\lambda 1.0694$~$\mu$m  
in the NIR at higher velocity
($\sim 10,000$~\kmps). 
Alternative identifications for the optical
feature are due to lines of  $\ion{Co}{ii}$ and $\ion{Si}{ii}$. 
 
  We note that in  PDDs there can be extra energy left over from
  shock heating in 
  the unburned regions 
with may lead to higher temperatures than in classical DD models
\citep{dessart14b}. Some energy from interaction my be also lead to
higher temperatures and hotter spectra \citep{Gerardy04}. Mixing of
radioactive material 
through the C+O region is
disfavored as we have already discussed above.

\subsection{Late phase spectroscopy}
\label{nebspectra}

 Independent of the details of the explosion scenario the inner
  regions of SNe~Ia ejecta are quite similar. Modulations due to
  variations in PDD models do not impact the bulk of the inner regions
  of the ejecta for models that produce the same amount of \nni
\citep{hkm95,hk96,dessart14b}. 
As discussed above we can probe the $^{56}$Ni distribution using our
reference model.

\subsubsection{Methods}

 Calculations of explosions, light curves, and spectra are performed 
 using the HYDro RAdiation transport (HYDRA)   code
\citep{h90,h95,hoeflich03,h09}. HYDRA solves the hydrodynamics using the
explicit Piecewise Parabolic Method  \citep[PPM][]{CW_PPM84}, 
detailed nuclear and atomic networks \citep{Cy10,k94d,h95,seaton05},
radiation transport including
$\gamma$-rays
and positrons by variable Eddington Tensor
solvers and Monte Carlo Methods  
\citep{mm84,stone1992,hoeflich03,h93,h09m,penney11}.  
For this study, forbidden line transitions have been 
updated using the atomic data in \citet{soma10a,soma10b}, \citet{brian_14J14}, and ``The Atomic
Line List V2.05B18'' at 
\url{http://www.pa.uky.edu/~peter/newpage/}
by Peter van Hoof.  
 For the overall evolution, we use simplified atomic models limited to about 2,000 --
 10,000 discrete levels using ``level merging'' or superlevels
 \citep{anderson89,DW93,HL95,schweitzer00,hillier03,hoeflich03,h09}.
 To compute NIR line profiles, we do not use superlevels but calculate the emission by
postprocessing of the background model \citep{hoeflich04,penney11,sadler12,h13}.
  Nevertheless, line ratios and excitation state depend 
sensitively on collisional de-excitation --- uncertainties in the
collisional cross sections have been and continue to be the main
uncertainies in determining the level populations
 \citep{axelrodphd,Bowers97,liu97a,Gerardy07}.

\subsubsection{Spectral Analysis}

All thermonuclear explosion models show centrally peaked density
profiles, therefore at late times, due to simple geometrical dilution
line formation should occur closer to the center at lower
  velocities yielding narrow line profiles. 
  In fact, if we see to the center the line profile
should peak at zero velocity, due to the central caustic.
However, due to the mass-radius relation of WDs, lower
  mass progenitors will have significantly lower central densities. If
  the central density becomes very low, such as in a sub-Chandrasekhar
  detonation or merger, than calcium will not be burned at low
  velocity, as noted above.

The low velocity Fe-group material seen in the late phase
  visual-wavelength spectra are a natural result of the PDD scenario,
  but can also be explained by pure 
  deflagration models that produce very low energy explosions
  \citep[see, for example][]{R02}, as well as by deflagrations leaving bound remnants \citep{kromer13,fink13}, and merger models \citep{Piersanti03,pakmor10}. 

Late phase spectra beyond 200 days are dominated by forbidden emission
lines which are powered by positrons originating from the $\beta^+$
decay of \nco. At this phase, hard $\gamma$~rays from the decay escape for all
classes of explosion scenarios \citep{hkm92}.  Up to about one year
after the explosion, the mean free path of positrons is small, leading
to  local energy deposition and excitation of levels
\citep{milne02,penney11,sadler12}, which traces the initial \nni
distribution. Thus, emission line profiles at optical and NIR
wavelengths traces the \nni distribution in velocity space, though blends may broaden the
features and affect the details of the line profiles,  particularly 
in the optical.  

NIR spectra at late epochs of normal SNe~Ia are dominated by blended
 forbidden lines of 
$[\ion{Fe}{i}]$, $[\ion{Fe}{ii}]$, $[\ion{Fe}{iii}]$, $[\ion{Co}{ii}]$ and
$[\ion{Co}{iii}]$ \citep{hoeflich04,motohara06}.  
As  normal SN~Ia evolve in
time, the strengths of their NIR spectral features change due to the
ionization state of the line forming regions, which is linked to the
ever decreasing decay of $^{56}$Co.  
During  this time the 1.5-1.7~$\mu$m region is dominated by a
multitude of blended  
forbidden $[\ion{Fe}{ii}]$ and $[\ion{Co}{iii}]$ lines, with major
contributions attributed to $[\ion{Fe}{ii}]$ $\lambda$$\lambda$1.53,
1.64, 1.67~$\mu$m, and $[\ion{Co}{iii}]$ $\lambda$1.55 and 
$\lambda$1.76~$\mu$m \citep{hoeflich04}. 
Eventually, the IR-catastrophe \citep{fransson94}
is expected to cause a redistribution of the energy deposition to the NIR
and beyond, however light curve observations of SNe~Ia constrain this possible
phase to much later epochs than considered in this paper
\citep{sollerman04,stritzinger07,Leloudas09,mccully14a}.

Fig.~\ref{fig:IR_line_descrip} shows  the comparison between
the  $+215$d optical (left panel) and  $+269$d NIR (right panel) spectra of 
SN~2012Z compared to the $+220$d reference model.
 As discussed above, the spectra are dominated by single and doubly ionized
lines of iron group elements, with almost all of the optical features 
formed from blends of various species. 
Many of the more prevalent observed features are reproduced by the
synthetic spectrum, however some discrepancies are apparent.
The  prominent $\sim$7400~\AA\ feature is well-matched by the 
synthetic spectrum. However, the observed feature is slighly
asymmetric on the blue wing, whereas the synthetic feature shows
multiple components formed by  
$[\ion{Ca}{ii}]$ $\lambda$7325.9, $[\ion{Fe}{ii}]$ $\lambda\lambda$7390.2,
7454.6 7639.6, with a significant 
contribution from $[\ion{Ni}{ii}]$ $\lambda\lambda$7379.9, 7413.6.
Here $[\ion{Ca}{ii}]$ contributes significantly to the line width of the blue wing.
The blue features are too strong in the synthetic spectrum, which is
likely due to inaccuracies in the  atomic data and the use of superlevels.
 
 Other notable  features are found to be formed by
$[\ion{Fe}{iii}]$ $\lambda\lambda$4659.5, 4703.0,4735.3, 4770.9,  
$[\ion{Fe}{iii}]$ $\lambda\lambda$4932.0, 5012.8, 
$[\ion{Fe}{ii}]$ $\lambda$5113.1, 
$[\ion{Fe}{ii}]$ $\lambda$5160.2, 
$[\ion{Fe}{ii}]$ $\lambda\lambda$5221.5, 5263.1,
$[\ion{Fe}{iii}]$ $\lambda$5272.0,  
$[\ion{Fe}{ii}]$  $\lambda\lambda$5263.1, 5298.3, 
$[\ion{Fe}{ii}]$  $\lambda\lambda$5335.1, 5377.9, 
$[\ion{Co}{iii}]$ $\lambda\lambda$5890.1, 5908.4 blended with 
 $[\ion{Fe}{ii}]$ $\lambda5872$,
$[\ion{Co}{iii}]$ $\lambda$6578.1,
$[\ion{Ni}{ii}]$ $\lambda$6668.6, and 
$[\ion{Co}{iii}]$ $\lambda$6855.4 blended with 
$[\ion{Fe}{ii}]$ $\lambda\lambda 6875.7, 6898.1$. 

 In addition, we see 
 $[\ion{Fe}{ii}]$ $\lambda$8619.3 and
$[\ion{Fe}{iii}]$ $\lambda\lambda$8840.6, 9036.0 though it appears 
slightly too low, most  likely due to discrepancies in the  atomic line data.
$[\ion{C}{i}]$ $\lambda$8729.5 may contributes
  to the observed feature at $\sim 8600$~\AA; however,  in the reference model, it forms
  at too high a velocity and is not the dominant transition. 
Another interesting feature present in both the models and SN~2012Z is the 
$[\ion{O}{i}]$  $\lambda\lambda$6302.0, 6302.0, 6393.5 triplet,
which appears rather weak.
The weak oxygen excitation expected in both DD and PDD
  models is due to the 
separation of $^{56}$Ni and the region of explosive oxygen burning
(that is where silicon and sulfur are formed), if
mixing in the PDD scenario
remains limited to a region smaller than the Si/S shell, as it appears
to be in SN~2012Z.

The NIR synthetic spectrum of Figure~\ref{fig:IR_line_descrip}  also shows
a host of $[\ion{Fe}{ii}]$ and $[\ion{Fe}{iii}]$ lines as seen in the
optical. Specifically showing features  at $0.95, 1.25, 1.5, 1.6, 1.8~\mu$m, including the strong feature at $1.25~\mu$m which is a blend of
$[\ion{Fe}{ii}]$, $[\ion{Fe}{iii}]$ and $[\ion{Ni}{ii}]$. 
An interesting feature located $\sim$~1.032~$\mu$m is attributed to a multiplet of $[\ion{S}{i}]$. 
Even moderate microscopic mixing of $^{56}$Ni would highly excite
the levels of $[\ion{S}{i}]$ producing a very prominent feature. Even higher
excitation, due to more mixing would produce a strong $[\ion{S}{ii}]$ feature at $1.02-1.04$~$\mu$m.
\footnote{Microscopic means on scales smaller than the  mean  free path for
positrons \citep{hoeflich02a,penney2014}.}

It is well established that the late-time NIR line profiles
of  $[\ion{Fe}{ii}]$ features trace the energy input by
radioactive decay \citep{hoeflich04,motohara06,maedanature10}.
 Because of the structure of the atomic levels, 
  the relatively unblended $[\ion{Fe}{ii}]$ 1.64~$\mu$m line is a particularly
  good tracer of the inner 
  region.
In a handful of the SNe~Ia observed several hundred days past maximum,
NIR emission lines have been found to exhibit pot-bellied
profiles.
 This observational characteristic is the hallmark of
$M_{\text{Ch}}$ WD progenitors that explode with high central
densities, of the order $\lesssim 10^9 $~\gcm
\citep{hoeflich04,motohara06,hoeflich06b,fesen07,Gerardy07,maedanature10,h13}.
At masses close to the $M_{\text{Ch}}$, central densities can become
sufficiently high that electron capture processes shift the NSE
distribution to stable isotopes of Fe/Co/Ni, leaving a \nni hole at
the center of the supernova ejecta, and thereby generate the
pot-bellied profiles.  This produces a central hole in \nni of about
3000~\kmps without mixing, very similar to SN~2003du and SN~2003hv
\citep{hoeflich04,motohara06}.  As pointed out in
Section~\ref{lateir}, the $[\ion{Fe}{ii}]$ $\lambda$1.64~$\mu$m feature
of SN~2012Z also exhibits a pot-bellied profile (see
Figure~\ref{LTnirscomp}), providing an additional indication that its
progenitor was at or near the $M_{\text{Ch}}$.

For clarity on this point, in Figure~\ref{nirFe2line}, we show the
effect of the $^{56}$Ni distribution on the late phase ($+$269d)
$[\ion{Fe}{ii}]$ $\lambda$1.64~$\mu$m line profile.  Here the observed
feature (black line) is compared to the model calculations of
5p0z22.16 at $+$220d for two cases: (1) a hole in the $^{56}$Ni
distribution of $\sim$3,000 km~s$^{-1}$ (blue line) and, (2) with the
$^{56}$Ni placed at the center of the ejecta (red line).  
A magnetic field of 10$^{6}$~G has been
assumed.  The blue line  is clearly pot-bellied, as in the
case of subluminous SNe~Ia at the same epoch, because some
$\approx$~6\% of all gamma rays excite the high density,
central region \citep{hoeflich91,hoeflich02a}.
  
The case for a $M_{\text{Ch}}$ explosion is quite strong, lower
initial WD masses would produce \nni confined to the central regions,
due to the lower densities in lower mass WDs.  
More centrally confined \nni would have longer diffusion times, thus
falling closer to the the light-curve brightness decline-rate relation and
it would not produce the observed spectra since the longer diffusion
times would lead to cooler (redder) light curves and spectra more
similar to fast decliners like SN~1991bg. Moreover, the persistent pot-bellied line profiles in the NIR require high densities only found in
$M_{\text{Ch}}$ explosions. 
  
 \section{Further Comparison Between SN~2005hk and SN~2012Z}
 \label{sec:compobj}
 In this section we further explore the commonalities between the
 observational properties of SNe~2005hk and 2012Z.  The former of
 these objects is a well-observed SN~Iax that has been previously
 documented to be similar to the prototypical Type~Iax
 SN~2002cx \citep{phillips07,sahu08}.  As previously highlighted, 
 SNe~2005hk and 2012Z exhibit comparable 
 decline-rate values (Figure~\ref{dm15vslum}), 
 color curves  (Figure~\ref{optcolors}), and peak luminosity (Figure~\ref{uvoir}).
 We now proceed to compare the shape of their light curves,
 and their optical and NIR spectra at various epochs.
 
 Plotted in Figure~\ref{lcscomp} is the comparison of the
 $uvw1$- and $ugriBV$-band light curves of both supernova plotted vs. $T(B)_{max}$,
 where the filtered light curves of SN~2005hk have been shifted to
 match the peak magnitudes of SN~2012Z.  Overall the shape of the
 optical light curves are exceedingly similar, with perhaps the only
 subtle difference being that SN~2012Z appears to rise to maximum
 brightness at a slightly faster rate. This is consistent with the \nni being distributed somewhat closer to the surface, 
 leading to shorter 
 diffusion times. 

 Figure~\ref{optspeccomp} is a comparison between the $-8$d, $+$2d and
 $+$21d visual-wavelength spectra of SN~2012Z, and similar epoch
 spectra of SN~2005hk \citep{phillips07}.  Overall the two objects are
 remarkably similar, exhibiting similar continua, and nearly identical
 spectral features at all epochs.  Comparing the expansion velocities
 of SN~2005hk \citep[see][their Figure 9]{phillips07} and SN~2012Z
 (see Figure~\ref{velocities}), shows that the latter consistently
 exhibits $v_{exp}$ values which are about 1500 km~s$^{-1}$ higher for
 all ions.  The similar magnitude $v_{exp}$ values between the two
 events is indicative that they had comparable explosion energies,
 while the broader line widths observed in SN~2012Z can be attributed
 to an optical depth effect caused by its $^{56}$Ni distribution
 extending to higher velocities. This suggests that the \nni was
 produced in different regions in SN~2005hk and SN~2012Z, implying
 that the explosion mechanism varied either due to some variation in a
 violent merger or more likely, that they both occurred due to a
PDD explosion \citep{khokhlov91b,khokhlov93,hk96}, where the mass of the
 pulsational shell varied leading to \nni production further-in in
 SN~2005hk and further-out in SN~2012Z. 
 Thus, we can think of the variation between SN~2005hk and SN~2012Z 
 as variations in the kinetic energy of the \nni, and not variations in the total kinetic energy of the explosion.

Figure~\ref{nirspeccomp} shows a comparison between the NIR
spectrum of SN~2012Z obtained at maximum (top panel) and three weeks
later (bottom panel) to similar epoch spectra of SN~2005hk
\citep{kromer13}.  At both epochs the spectra of the two objects are
very similar, with the main difference again being the moderately higher
blueshifts of the absorption features in SN~2012Z.  Another noticeable
difference observed in the bottom panel is that the strength of many
of the absorption features appears stronger in SN~2005hk, particularly
the $[\ion{Co}{ii}]$ lines located between $\approx$~1.6-1.8~$\mu$m. This
is due to the Fe-group elements in SN~2005hk being slightly more
confined in velocity space. In fact, all of the line profiles are
somewhat more confined in velocity space due to the more central
nickel distribution in SN~2005hk. 

\section{Discussion}
\label{sec:discussion}

From the findings presented in the previous sections it is clear that
SN~2012Z is among the brightest and most energetic SN~Iax yet
observed, further extending the parameter space of this peculiar class
of transients.  It is also apparent that objects populating the bright
end of the SN~Iax luminosity distribution exhibit exceedingly similar
observational parameters.  In Section~\ref{sectionuvoir} we derived
explosion parameters based on a number of underlying assumptions that
provide model fits to the UVOIR light curves of SNe~2005hk and 2012Z.
The model fits suggest $^{56}$Ni values ranging between 0.2-0.3
$M_{\sun}$, $E_{K}$ values of $\approx$ 10$^{51}$ erg, and 
ejecta masses consistent with $M_{\text{Ch}}$.

We have presented a possible explosion scenario that explains the
nature of at least some SNe~Iax. This involves a PDD explosion
 of a near $M_{\text{Ch}}$  WD . The production of
Fe-group elements occurs in the deflagration phase (in distinction
to normal SNe~Ia where the Fe-group elements are primarily produced
in the detonation phase, leading to a more layered or radially 
stratified structure).
Our scenario explains both the photospheric velocities of the IMEs, as
well as the high densities required by the late phase 
NIR spectra. The production of \nni closer to the photosphere also
simply explains the blue colors and hot spectra.  
From our reference model and the series of PDD models of  \citet{hkm95} we
would find the total  mass of \nni to be  0.15 and 0.2 \msol,
respectively.

SN~2012Z provides a uniquely large data set which allows for a
detailed analysis covering a wide range of observables unparalleled
even for typical SNe~Ia.  The nature of SN~2012Z is a
theoretical challenge, because several key physics aspects in thermonuclear
explosions are not well understood, such as: the progenitor evolution;
the thermonuclear runaway; rotation; and so on, making fully self
consistent models a goal for the future. These problems lead to the wide
variety of models suggested for SNe~Iax.  

Using both analytic and
numerical models as benchmarks, we were able to develop constraints
for SN~2012Z based on multiple observables.  The light curves 
suggest a massive, close to $M_{\text{Ch}}$, WD as the progenitor.
From
the peak brightness we may estimate the mass to be within $\sim 0.2~M_\odot$
of $M_{\text{Ch}}$. This view is
supported by the pot-bellied, late time NIR line profiles that can be
understood by high-density burning taking place at densities in excess 
of $10^{9}$~\gcm, which puts the progenitor within about $0.02~M_\odot $ 
of the $M_{\text{Ch}}$. 
The other, less stringent constraint, $M > 1.2 M_\odot$,
is from the lack of Ca II (and other IMEs) at low velocities, which even at solar
abundance, produces very strong features in the optical and NIR. Thus,
solar abundance primordial calcium must have been burned at low
velocities. 

Light curves are mostly determined by diffusion time
scales, that is, their appearance is dictated by the nature of the
innermost regions. The light curve rise time is determined by the
nickel distribution. Even with $M_{\text{Ch}}$ progenitors, fast rise times and
slow declines can be reproduced \citep{baron12}. Within this model
elements produced in the deflagration will be mixed, while those
produced in the detonation will have a layered structure. 
Note that the  detonation  running through the material previously burned  in the deflagration will reproduce the layered structure.

PDDs are similar to deflagrations, but have a detonation phase which
produces a layered  structure without requiring a bound
remnant. Because PDD models may vary  in the ratio of
  energy released to binding energy \citep{quimby07}
   they produce a wide range of
\nni masses. 
 Also since the shell mass may vary, a range of unburned C+O in the
 outer layers  can also be in PDDs. 

The similarity between SN~2012Z and SN~2005hk suggests
very similar properties of the exploding WDs. However, SN~2012Z has
undergone relatively enhanced mixing to bring Fe-group elements to the
base of the incomplete oxygen burning (Si/S) region.
All multidimensional models predict
mixing by RT instabilities which produce an almost homogeneous mixing
of the layers of IMEs with those below. In
SN~2012Z, we see a layered structure, the signature
of detonations, for calcium, silicon, and magnesium.  This suggests that these elements
are produced in a 
detonation phase after the mixing has occurred and the majority of Fe-group
elements have been synthesized. RT instabilities in the deflagration phase
seem to be a stable feature from the hydrodynamical view making
deflagration models much less favorable as compared to models 
that result in the
layered structure observed in SNe~Ia. 
Detonation fronts in a WD will
produce IMEs in a wide velocity range due to the
smooth density structure and thus, slowly varying burning conditions.
An exception to this picture can be found in the PDD  models,
where  low densities in an shell prevent
burning to the Fe-group by a detonation, and limit the expansion
velocities of the 
burning products. For SN~2012Z, we need  a shell mass of about
$0.05$-$0.12~M_{\odot}$ to be 
consistent with the low range of velocities observed in IMEs of SN~2012Z.
 We also need little or no production of \nni
during the detonation phase.
PDDs model will not show  signatures of narrow (low velocity) oxygen features 
at late phases because oxygen is consumed during the detonation
 phase. Moreover, as the expansion velocities of SNe~Iax are roughly half of what 
 is observed in normal SNe~Ia,  densities are kept sufficiently high 
 to suppress spectra dominated by forbidden  lines of Fe-group elements
 out to eight times 
 later than observed in SNe~Ia, or roughly two to three years past explosion.

Because most of the \nni production is during the deflagration
phase,  we expect a wide variation in the $\Delta m_{15}$ relation
compared to normal SNe~Ia in which most of the \nni is produced
during the detonation phase. This is due to the fact that the amount
of \nni  produced and its spatial distribution will depend sensitively
on the mixing during the deflagration phase, rather than the smooth
layered nickel distribution produced in the detonation phase.

PDD models synthesize abundances of \nni  that range from near zero to 0.8~\msol,
and therefore offer the flexibility to produce both bright and dim events.
While our proposed model seems to do a reasonable job of explaining
the moderately bright SN~2005hk and SN~2012Z, and perhaps also 
the low luminosity 
SN~2010ae, we should stress the limitations of our analysis. It is not
yet clear whether  all SNe~Iax are a homogeneous class, or like
SNe~Ia, will show wide variations. 
 Our analysis here has only been a qualitative description
  of our scenario.
  To understand how widely applicable it is
  to SNe~Iax as a class a detailed parameter study with radiative
  transfer needs to be compared to large datasets of individual SNe~Iax.

 A common problem with both the PDD model and models with C$+$O shells, depending on the mass and the amplitude of the explosion,
may be the optical depth of the shell during early times which could lead to high velocity calcium and other background metal lines.
 Accretion from a helium or low metallicity star may
reduce this problem due to the low opacity of both unexcited helium
(since gamma-rays are shielded from the surface). If there is a
significant unburned C+O shell then CO molecules may form 
  which would provide an important coolant and should
be visible in the mid infrared \citep{hkm95}. 

Pure deflagration models have been suggested as a
candidate scenario for SNe~Iax since they can produce a wide range
of  kinetic energies, and peak brightnesses
\citep{phillips07,sahu08}. 
Depending on the initial conditions, pure deflagrations may or may not leave a bound
remnant  \citep{R02,fink13}.  
 Recently
a suite of model calculations has been presented, especially 
designed for SNe~Iax,
that consist of
C$+$O WDs that undergo deflagration driven
disruption  leaving  bound
remnants \citep{jordan12,fink13,kromer13}. 

\citet{kromer13} presented a pure deflagration model 
 leading to a bound remnant of nearly 1~\msol, which  produces
enough $^{56}$Ni to account for the peak brightness of 
SN~2005hk. Their model does a reasonable job of 
fitting the  photospheric phase optical spectral around maximum light,
and the fast rise and slow late-time decline  of the light
curve of SN~2005hk. In
this model the remnant serves to explain the absence of late-time
oxygen lines that are produced by the incomplete burning and mixing
inherent in 
pure deflagration models. 

Any model that involves a pure deflagration will give a non-layered
structure with mixing of unburned material, IMEs,
and Fe-group elements throughout the entire ejecta. In particular,
this would produce low-velocity oxygen lines at late times
\citep{kozma05}. The
model of \citet{kromer13} avoids the appearance of low
velocity unburned material by  
leaving the central regions behind in the bound remnant, 
while not producing a radially stratified ejecta.
Late-time spectra of SN~2012Z (see Figure~\ref{nebcomp}) show shell like line profiles which may be produced
in the bound remnant scenario,
but the low mass may lead to the formation of
true shell-like profiles, that is the formation of horned emission
lines.
It remains for future calculations to determine if the bound remnant model is able to produce disruptions that synthesize the small amounts of $^{56}$Ni
measured in the 2008ha/2010ae-like SNe~Iax. 
 In the same context, the 
  PDD models that are found in the literature have also not
  been extended to the dimmest extent in order to 
  be compared with
  the observations of the faintest members of the  SNe~Iax class.

 In many respects, other models invoking detonations, such as dynamical merging
  scenarios and helium detonations can not be ruled out for 
SNe~Iax as they may produce similar configurations as PDD models 
\citep{benz90,livio89,loren09,pakmor10,shen10,ruiter14}.
 However, they may also 
be somewhat disfavored due the lack of polarization observed in SNe~Ia
and SNe~Iax 
\citep{chornock06,maund10,patat12}. 
Additionally the late phase spectra of
SN~2012Z, seem to disfavor merger models due to lower densities in the
progenitors which leads to \nni all the way to the center (see
Section~\ref{lateir}). 

Characteristics in favor of PDD-like models come from  the
advantages of combining attributes of deflagration and detonation models. 
 As discussed above, the late phase spectra of SN~2012Z show evidence for burning at high densities. 

The observational properties described in this paper for SN~2012Z 
 have also not been probed for the
majority of SNe~Iax, due to incomplete data sets. 
We note that combining the observed characteristics  from different SNe~Iax
could lead to wrong conclusions.
Therefore we cannot exclude that a wide variety of scenarios 
may be realized within the SNe~Iax class. 

Understanding the nature of SNe~Iax will also come from determinations
about the progenitor system in individual cases.
\citet{mccully14b} have observed the progenitor system of SN~2012Z
with HST via serendipitous measurements of the host
galaxy. They find a stellar source for the progenitor, S1, which is
luminous and blue. Examining the color magnitude diagram for S1, it is found
 that it is consistent with a massive blue supergiant, which is a
very unlikely progenitor for SN~2012Z in any scenario. They conclude
that they may be seeing the donor companion star, likely a $\sim
2~M_{\odot}$ helium star. 
 While the  current observations are not completely aligned with our 
expectations of a hot disk as the source of
S1, it is possible that the environment is contaminated enough that at
least some of the light from S1 is indeed from the hot accretion
disk. Nevertheless, the observed progenitor fits in with a single
degenerate explosion which would be consistent with our proposed
scenario. More will be learned from future observations of the
putative progenitor when the light from the supernova has faded.

 In conclusion, we stress that future efforts to model the SN~Iax class 
should aim to provide robust radiative
transfer calculations in order to compare to the detailed optical and
NIR spectroscopic sequences now being routinely obtained at \textit{both}
early and late phases. We have shown that there are constraints that
models must obey, but quantitative comparison with the full SNe~Iax
class  requires  a wide parameter 
study of the variety of mixing and pulsation properties. Such a study is
 underway, and should provide confirmation or not of whether PDD 
models may indeed provide a viable explosion model for SNe~Iax.

\begin{acknowledgements}
  We thank M.~Tanaka and V.~Stanishev for providing access to their
  published spectra of SN~2005hk.
    A special thanks to L.~W.~Hsiao for providing assistance in some of
  the observations presented in this study, as well as to the Las
  Campanas technical staff for their continued support over the years.
  M.~D. Stritzinger, E. Hsiao, and C. Contreras gratefully acknowledge
  generous support provided by the Danish Agency for Science and
  Technology and Innovation realized through a Sapere Aude Level 2
  grant.  M.~D. Stritzinger, F. Taddia and S. Valenti acknowledge
  funding provided by the Instrument Center for Danish Astrophysics
  (IDA).  We would like to express our thanks to Peter van Hoof for
  creating the Atomic Line List V2.05B18 at 
  \url{http://www.pa.uky.edu/~peter/newpage/}.
  The CSP is  supported by the NSF under grants AST--0306969,
  AST--0607438 and AST--1008343. This work was also supported by the
  NSF to P. Hoeflich through grants AST--22111 and AST--23432. 
  E. Baron was supported in part by NSF grant AST-0707704. 
   S.~Benetti, is partially supported
  by the PRIN-INAF 2011 with the project ``Transient Universe: from
  ESO Large to PESSTO". 
  G.~Pignata acknowledges support provided by
  the Millennium Institute of Astrophysics (MAS) through grant
  IC120009 of the Programa Iniciativa Cientifica Milenio del
  Ministerio de Economia, Fomento y Turismo de Chile. 
  This research has made use of the NASA/IPAC Extragalactic Database (NED),
  which is operated by the Jet Propulsion Laboratory, California
  Institute of Technology, under contract with the National
  Aeronautics and Space Administration.
\end{acknowledgements}

\bibliographystyle{aa}

\clearpage
\begin{figure}[h]
\centering
\includegraphics[width=4.6in]{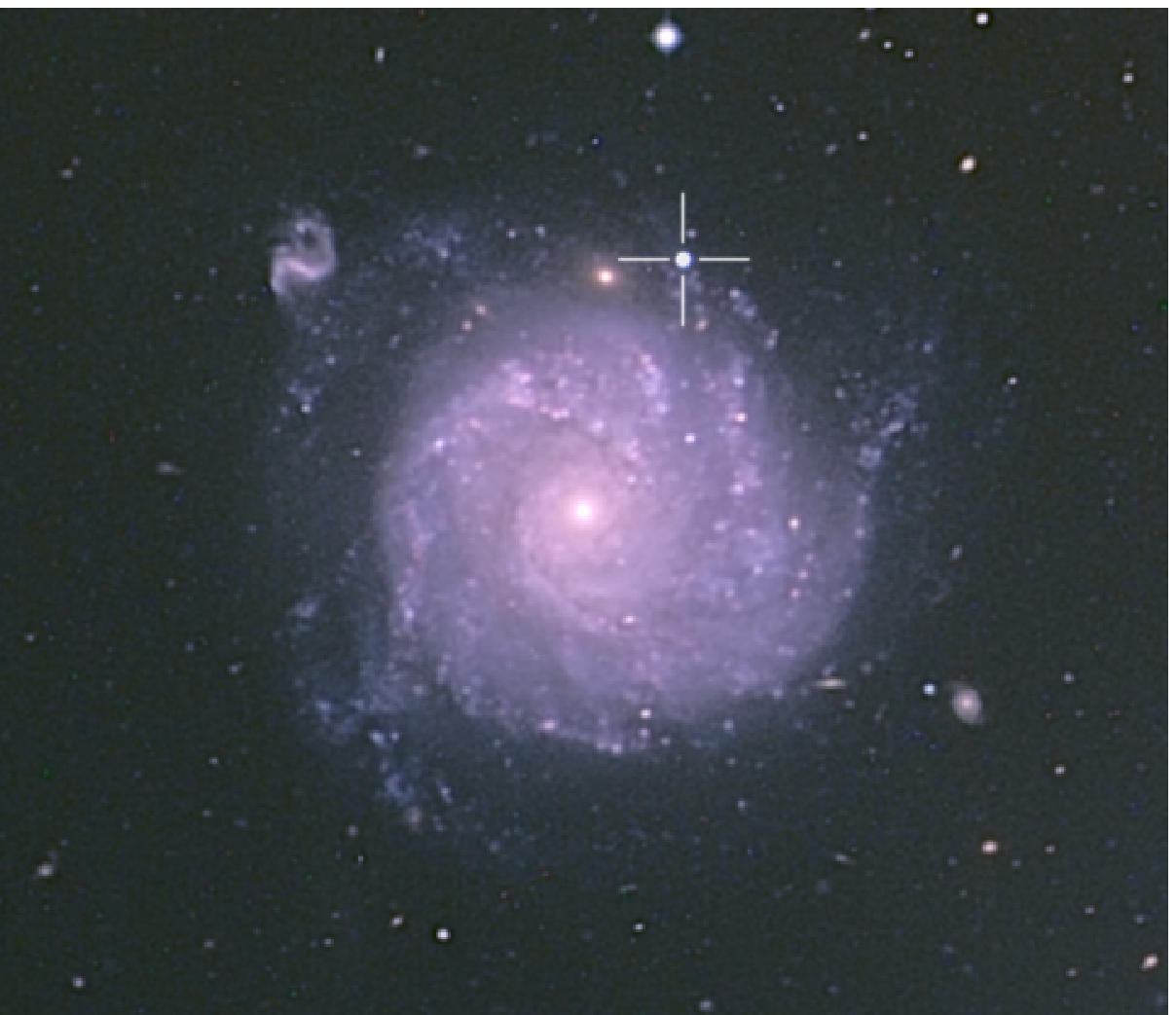}
\caption[]{Composite image of the SA(s)bc galaxy NGC~1309, with the
  position of SN~2012Z indicated. North is up and East is left.\label{FC}}
\end{figure}

\clearpage
\begin{figure}[h]
\centering
\includegraphics[width=5.6in]{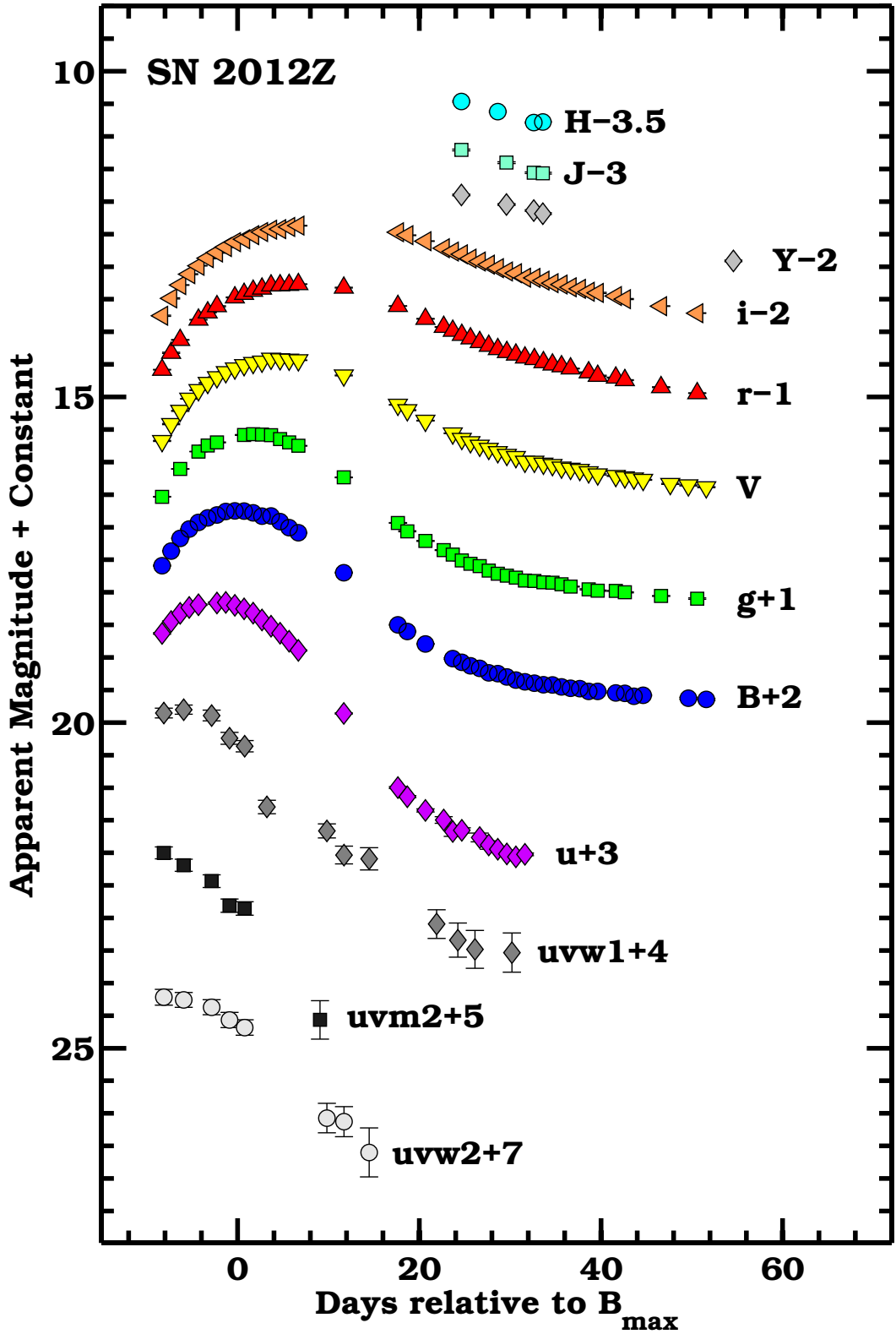}
\caption[]{UV, optical and NIR light curves of
  SN~2012Z.\label{earlylcs}}
\end{figure}

\clearpage
\begin{figure}[h]
\centering
\includegraphics[width=5.6in]{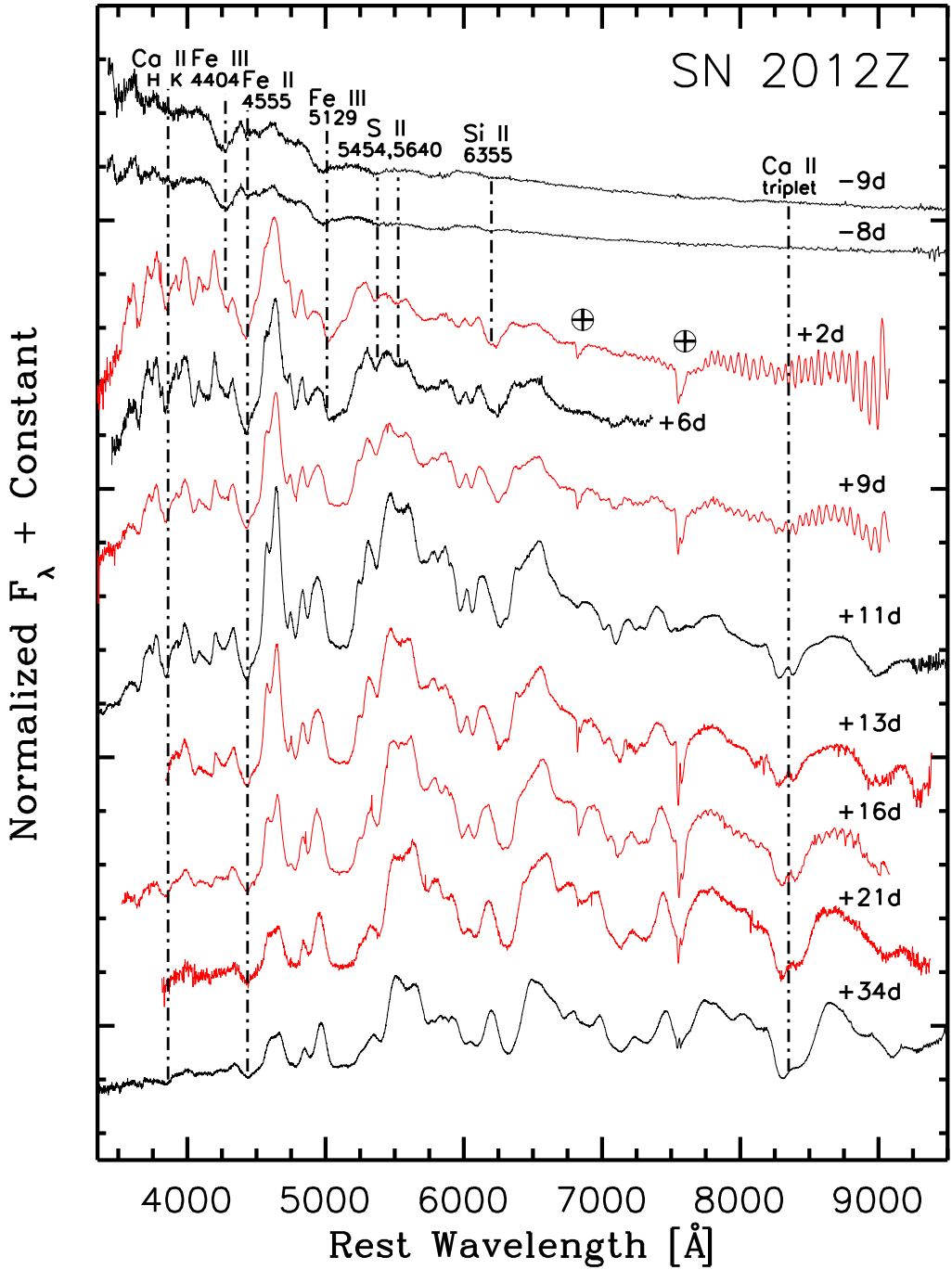}
\caption[]{Early phase, visual-wavelength spectroscopy of SN~2012Z
  plotted in the rest frame of the host galaxy.  A subset of spectra
  from \citet{foley13} are plotted in black, and our previously
  unpublished spectra are shown in red. Phase relative to $T(B)_{max}$
  is listed to the right of each spectrum. Telluric features are
  indicated with an Earth symbol. Note that spectra obtained on $+$2d,
  $+$9d, and $+$16d suffer from moderate fringing. \label{optspec}}
\end{figure}

\clearpage
\begin{figure}[h]
\centering
\includegraphics[width=5.6in]{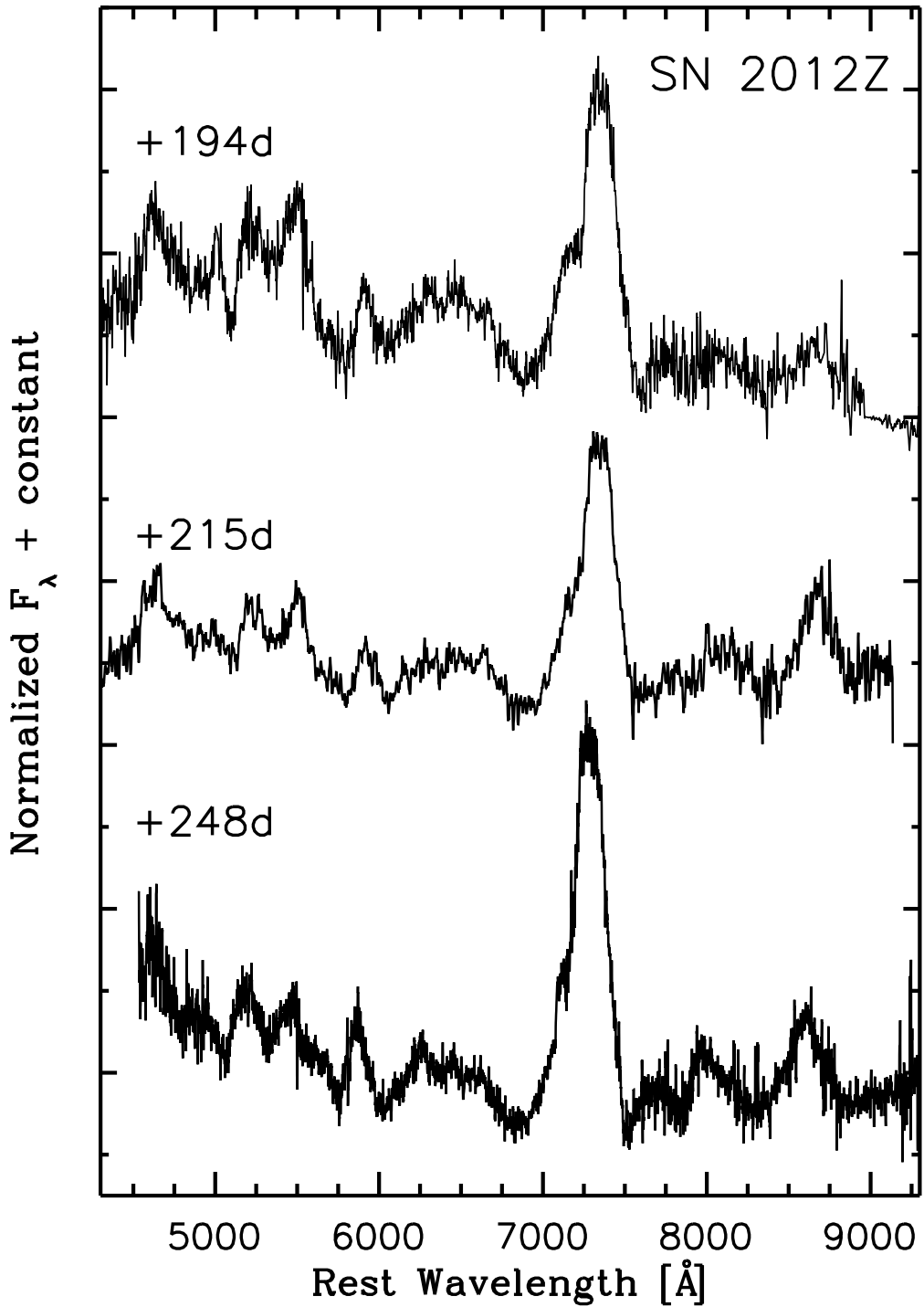}
\caption[]{Late phase, visual-wavelength spectra of SN~2012Z obtained
  on $+$194d, $+$215d and $+$248d. Each spectrum has been slightly
  smoothed for presentation purposes. \label{nebopt}}
\end{figure}

\clearpage
\begin{figure}[h]
\centering
\includegraphics[width=5.6in]{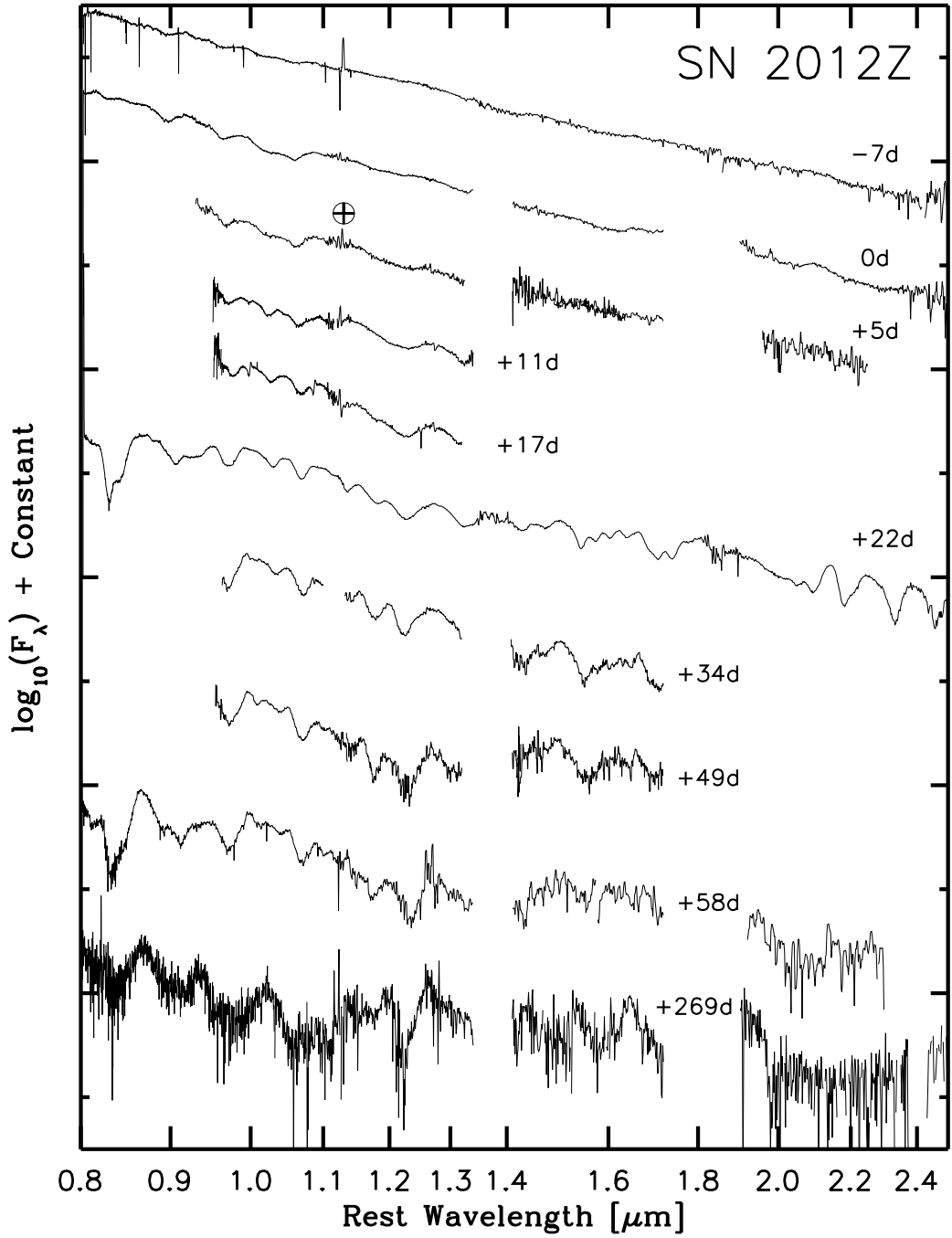}
\caption[]{NIR-wavelength spectroscopy of SN~2012Z.
Phase relative to $T(B)_{max}$ is listed to the right of each spectrum.
\label{nirspec}}
\end{figure}

\clearpage
\begin{figure}[h]
 \centering
\includegraphics[width=5.6in]{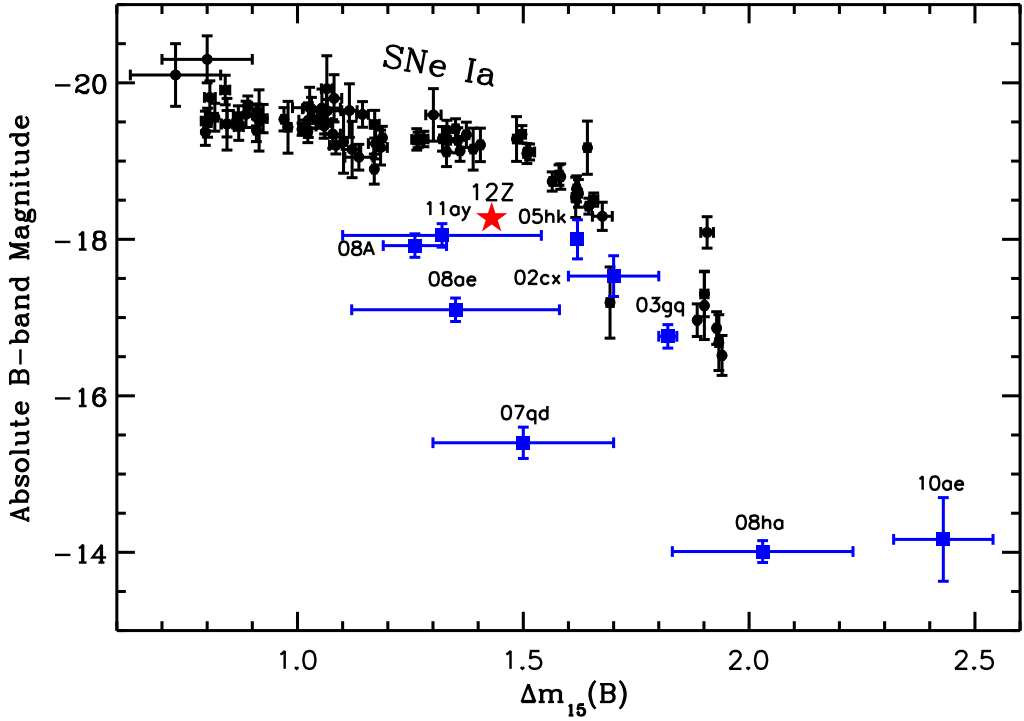}
\caption[]{$M_{B}$ plotted vs. $\Delta$m$_{15}(B)$ for a sample of  
CSP SNe~Ia  (black  dots), a handful of  SNe~Iax (blue  squares) that span their full range in luminosity, and SN~2012Z (red star). 
Note the error bars associated with SN~2012Z are smaller than its symbol size.
The comparison SNe~Iax plotted are
SN~2002cx \citep{li03,phillips07},
SN~2003gq \citep[see][and references therein]{foley13},
 SN~2005hk \citep{phillips07}, 
 SN~2007qd  \citep{mcclelland10}, 
 SN~2008A \citep[see][and references therein]{foley13},
 SN~2008ge \citep{foley13},
 SN~2008ha \citep{stritzinger14}, 
 SN~2010ae \citep{stritzinger14},
 and 
 SN~2011ay (Brown, private communication).
 \label{dm15vslum}}
\end{figure}

\clearpage
\begin{figure}[h]
\centering
\includegraphics[width=5.6in]{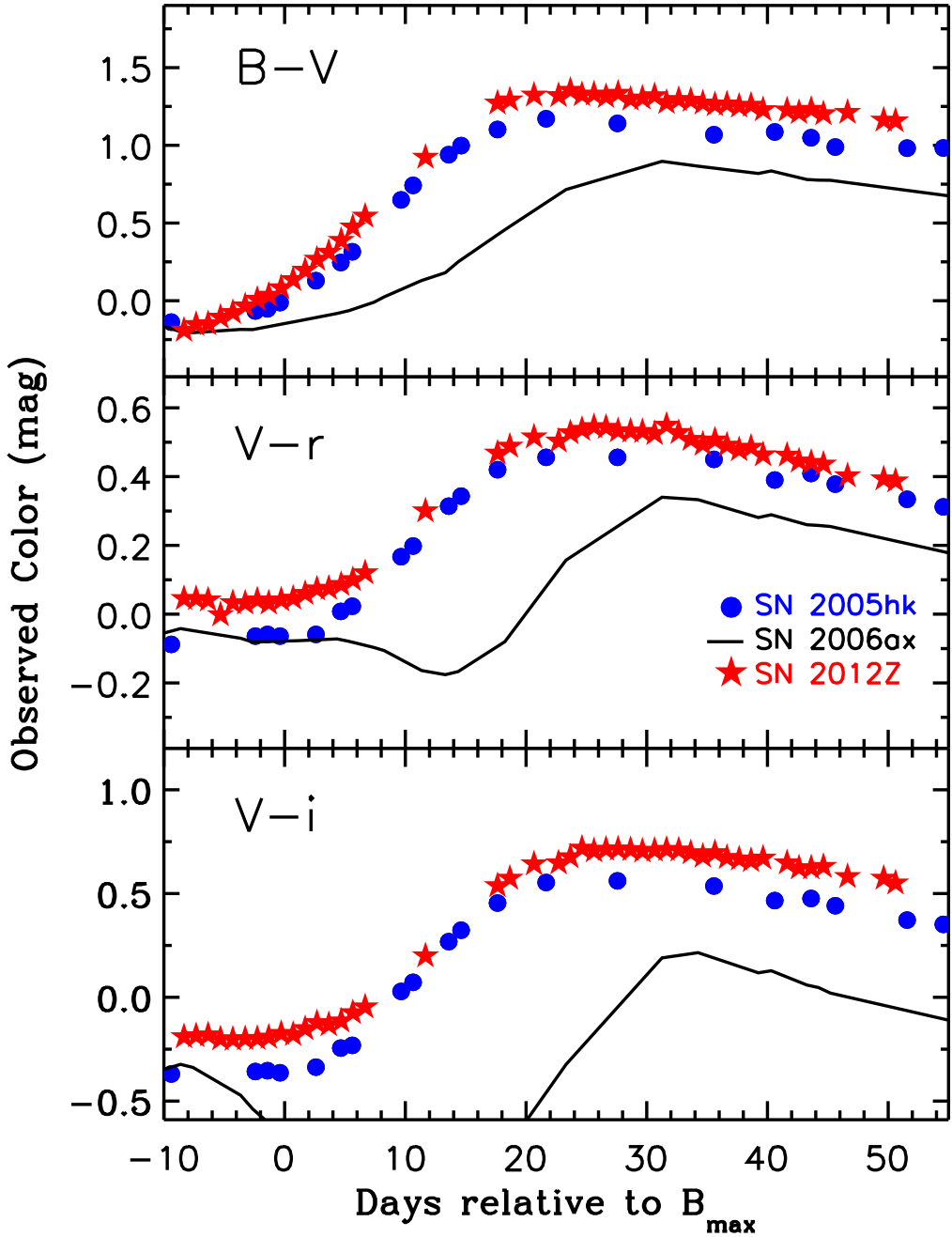}
\caption[]{$(B-V)$, $(V-r)$, and $(V-i)$ color evolution of SN~2012Z
  (red stars), 
 compared to the Type~Iax SN~2005hk (blue dots) and the un-reddened, normal
 Type~Ia SN~2006ax (solid line). The photometry of SN~2006ax 
 is taken from \citet{contreras10}. 
 The color curves of SNe~2005hk and 2012Z have been corrected for  Milky Way and host extinction adopting $E(B-V)_{tot} = 0.112$ mag \citep{chornock06,phillips07} 
 and $E(B-V)_{tot} = 0.105$ mag, respectively,
 while the color curves of SN~2006ax have been corrected for Milky Way reddening.
\label{optcolors}}
\end{figure}

\clearpage
\begin{figure}[h]
\centering
\includegraphics[width=5.6in]{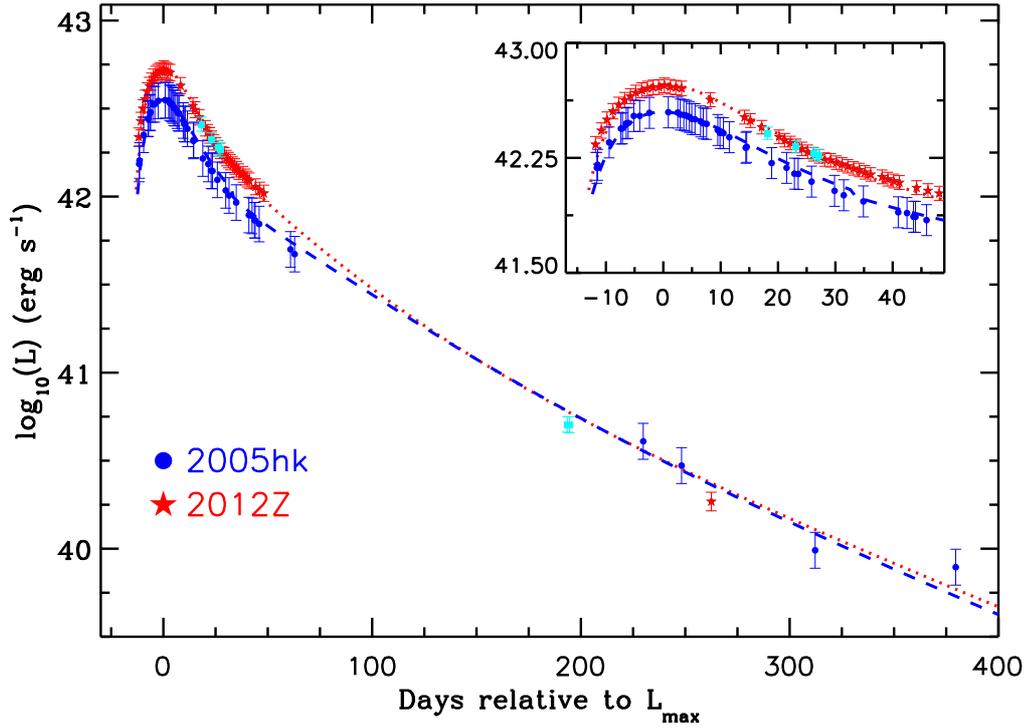}
\caption[]{UVOIR light curves of SNe~2005hk and 2012Z, over-plotted 
by model fits based on Arnett's equations. The model fits  provide
estimates of the $M_{Ni}$, $M_{ej}$ and $E_K$ (see
Section~\ref{sectionuvoir}).  
Also plotted as filled cyan squares are UVOIR flux points of SN~2012Z
computed using its optical and NIR photometry. These points are
in excellent agreement with the UVOIR flux points of SN~2012Z, as
computed using the NIR flux contribution estimated from SN~2005hk.  
Note that the error bars are appreciably larger for SN~2005hk, due to
its larger uncertainty in distance. 
\label{uvoir}}
\end{figure} 

\clearpage
\begin{figure}[h]
\centering
\includegraphics[width=5.6in]{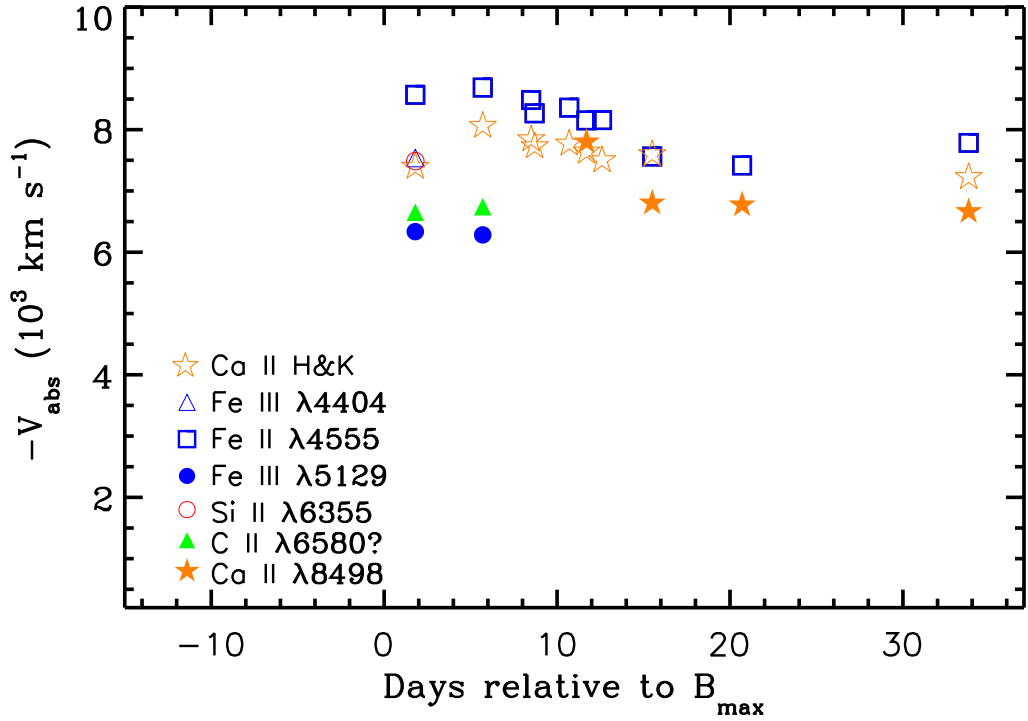}
\caption[]{Evolution of the Doppler velocity at maximum absorption 
of several prominent ions in the spectra of SN~2012Z. The
$\ion{C}{ii}$ feature is
not prominent in the observed spectra and the line
identification is suspect. Hence,  the inferred velocity
is somewhat questionable 
(see the text for more discussion). 
\label{velocities}}
\end{figure}

\clearpage
\begin{figure}[h]
\centering
\includegraphics[width=5.6in]{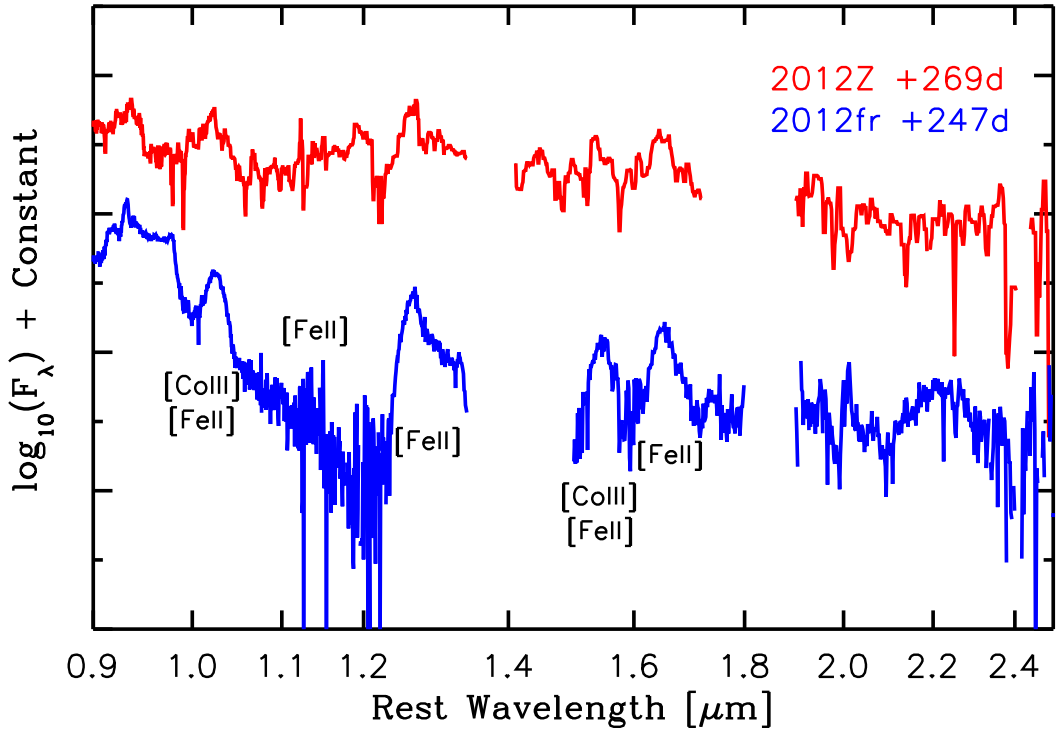}
\caption[]{Comparison of similar epoch late phase
NIR spectrum of the normal Type~Ia SN~2012fr and the 
Type~Iax SN~2012Z.\label{LTnirscomp}}
\end{figure}

\clearpage
\begin{figure}[h]
\centering
\includegraphics[width=5.6in]{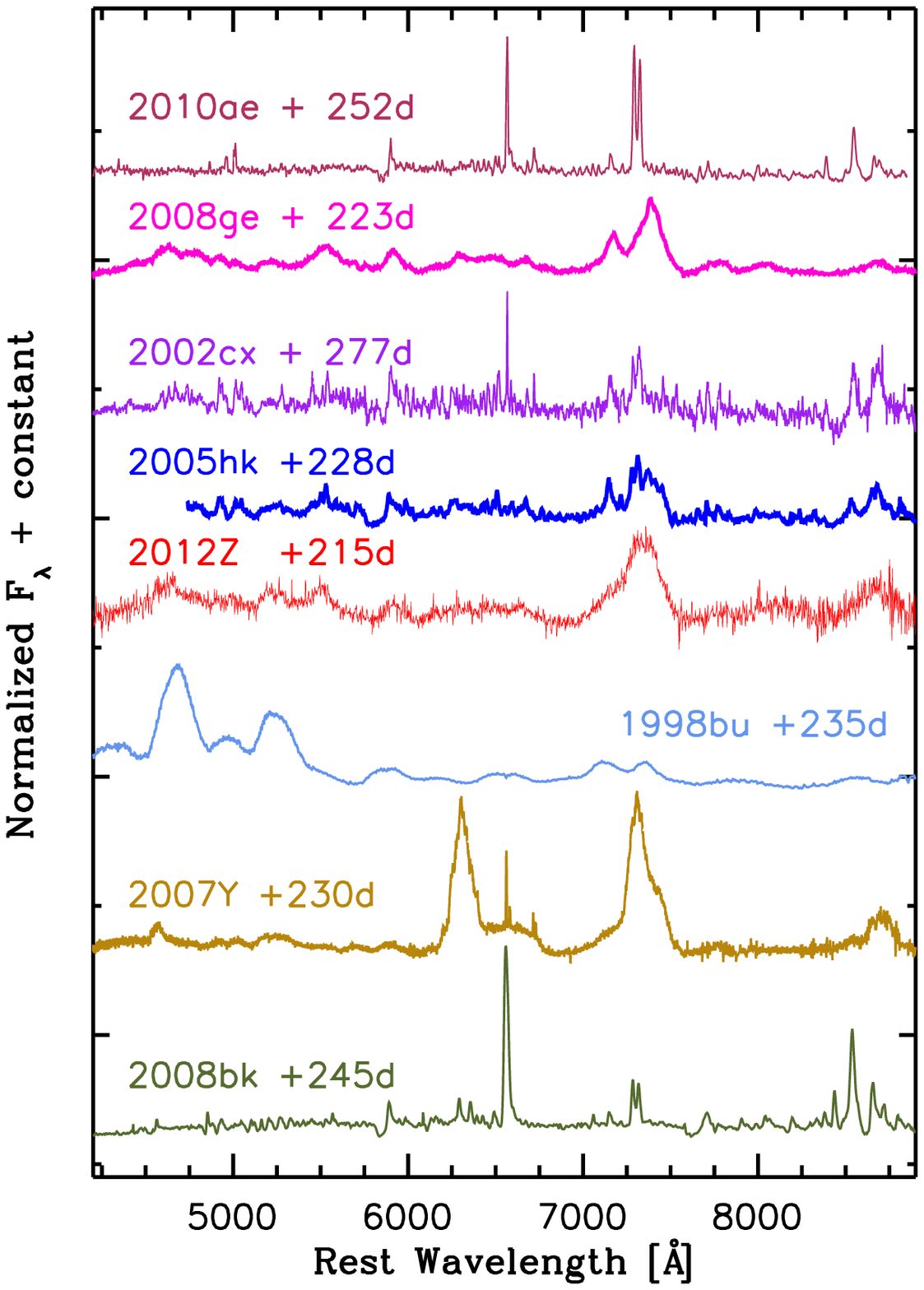}
\caption[]{Comparison of late phase visual-wavelength spectra of 
a number of  SNe~Iax and other SN types. 
The first 5 spectra, sorted    
from faintest to brightest (top to bottom), are the Type~Iax 
SNe~2010ae \citep{stritzinger14}, 
2008ge \citep{foley10},
2002cx \citep{jha06},
2005hk \citep{sahu08},
and 
2012Z.
Also shown are similar epoch spectra of the normal
Type~Ia SN~1998bu,
 \citep{silverman13},
the Type~Ib SN~2007Y \citep{stritzinger09},
and the subluminous Type~IIP SN~2008bk \citep{stritzinger14}.  Note
in SN~2008bk Balmer lines are associated with emission from the SN ejecta, rather
than with the narrower lines associated with host nebular lines, as is the case for SNe 2002cx and ~2010ae.}
\label{nebcomp}
\end{figure}

\clearpage
\begin{figure}[h]
\centering
\includegraphics[angle=0,width=6.6in]{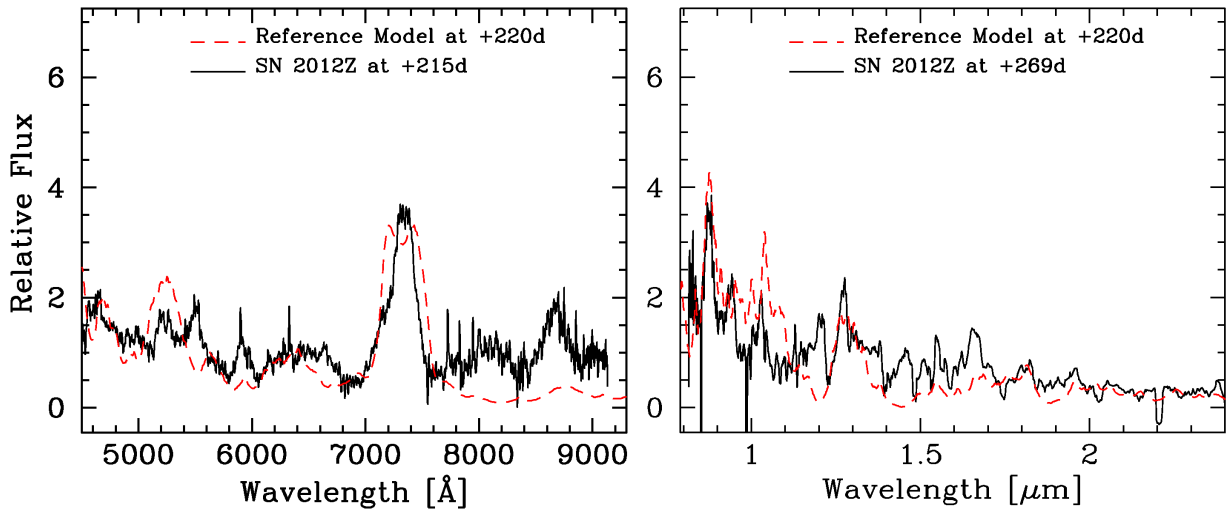}
\caption[]{Late phase optical (right) and NIR (right) spectrum of SN~2012Z compared to the synthetic spectrum computed from the reference model.\label{fig:IR_line_descrip}}
\end{figure}

\clearpage
\begin{figure}[h]
\centering
\includegraphics[width=6.6in]{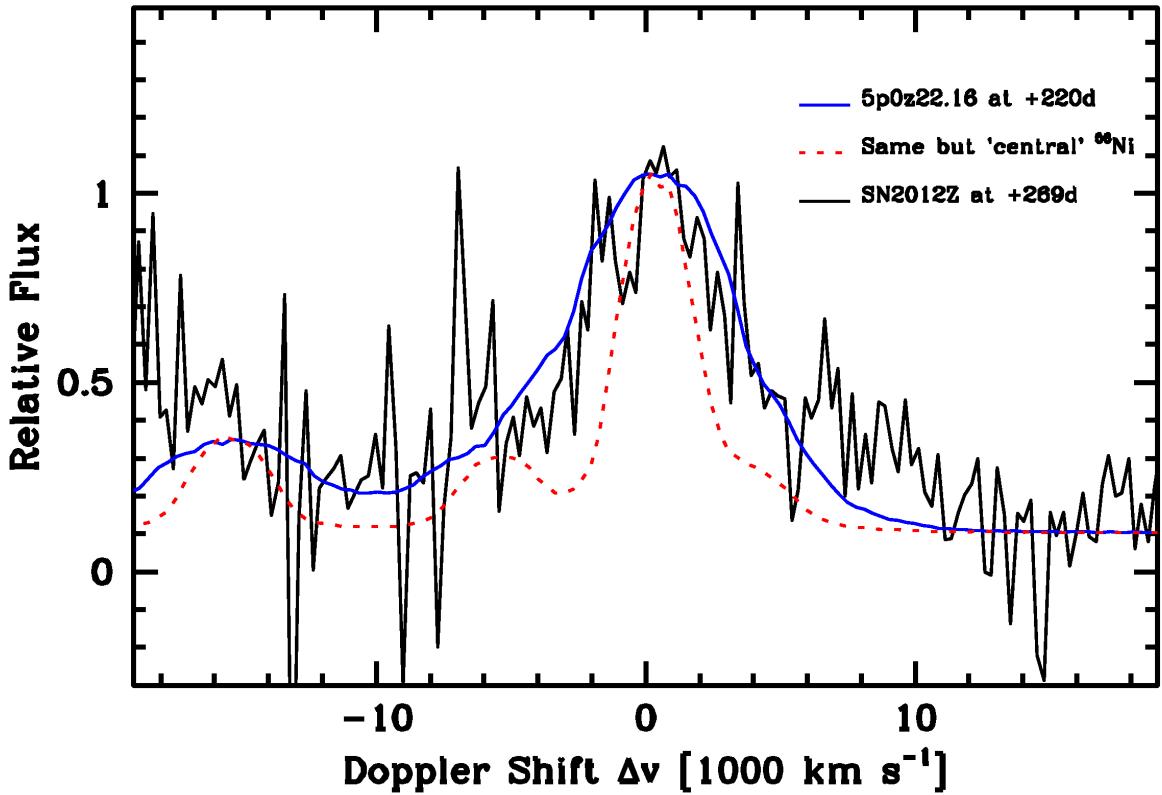}
\caption[]{Effect of the $^{56}$Ni distribution upon the late phase 
$[\ion{Fe}{ii}]$ $\lambda$1.64~$\mu$m feature.
Comparison of this feature from the $+$269d spectrum of SN~2012Z (black) to the expected line profile computed from the 5p0z22.16 model at $+$220d for two different $^{56}$Ni distribution. 
Plotted in blue is the synthetic spectra for the `original' model which has a `hole' in the $^{56}$Ni distribution of $\approx$ 3000~km~s$^{-1}$, while 
in red is the synthetic spectrum with $^{56}$Ni concentrated in the center.\label{nirFe2line}}
\end{figure}

\clearpage
\begin{figure}[h]
\centering
\includegraphics[width=5.6in]{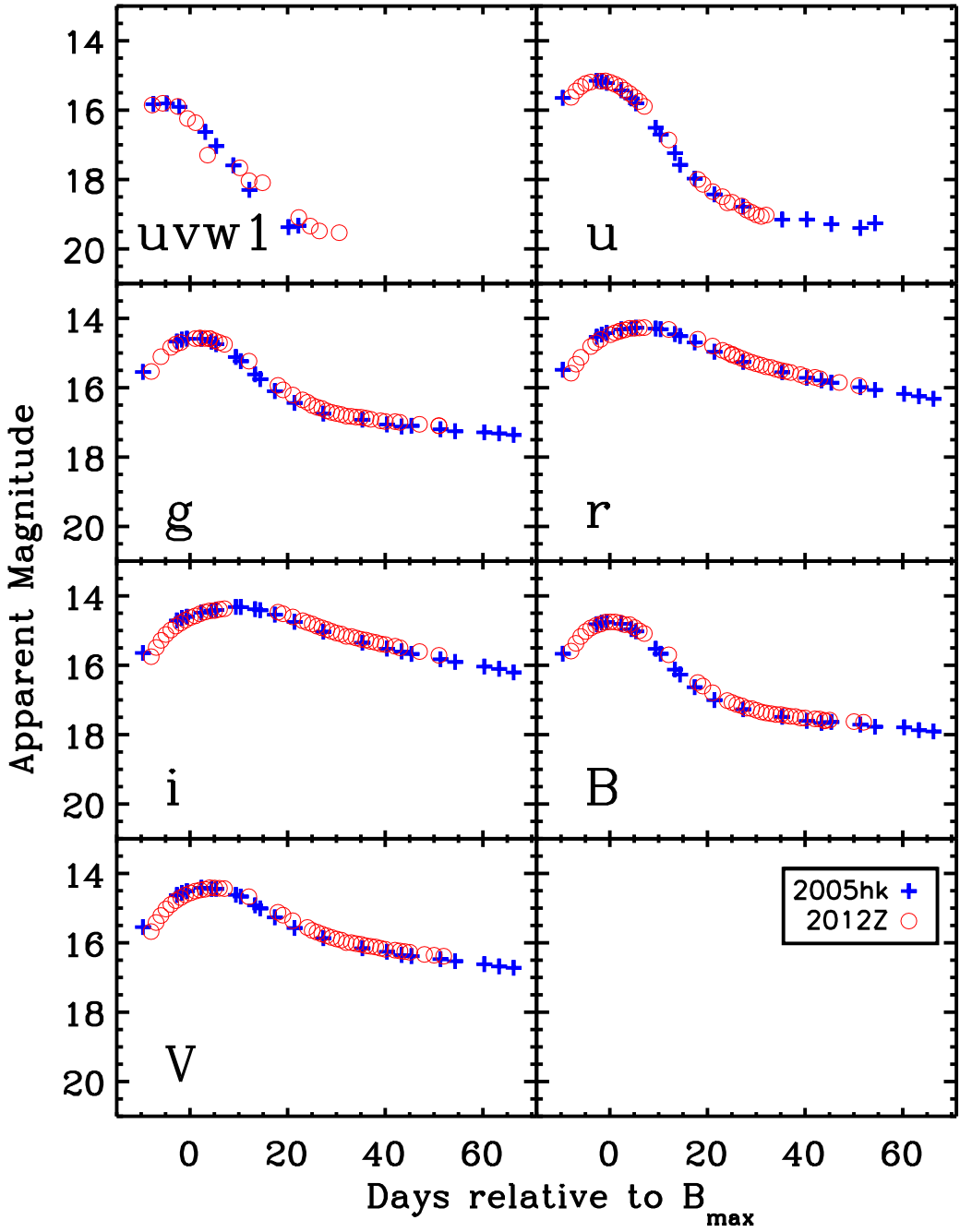}
\caption[]{Comparison of $uvw1$- and $ugriBV$-band light curves of SNe~2005hk
  (black plus symbols) and 2012Z (red circles). The light curves of
  SN~2005hk have been shifted to match the peak magnitudes of
  SN~2012Z. 
\label{lcscomp}}
\end{figure}

\clearpage
\begin{figure}[h]
\centering
\includegraphics[width=5.6in]{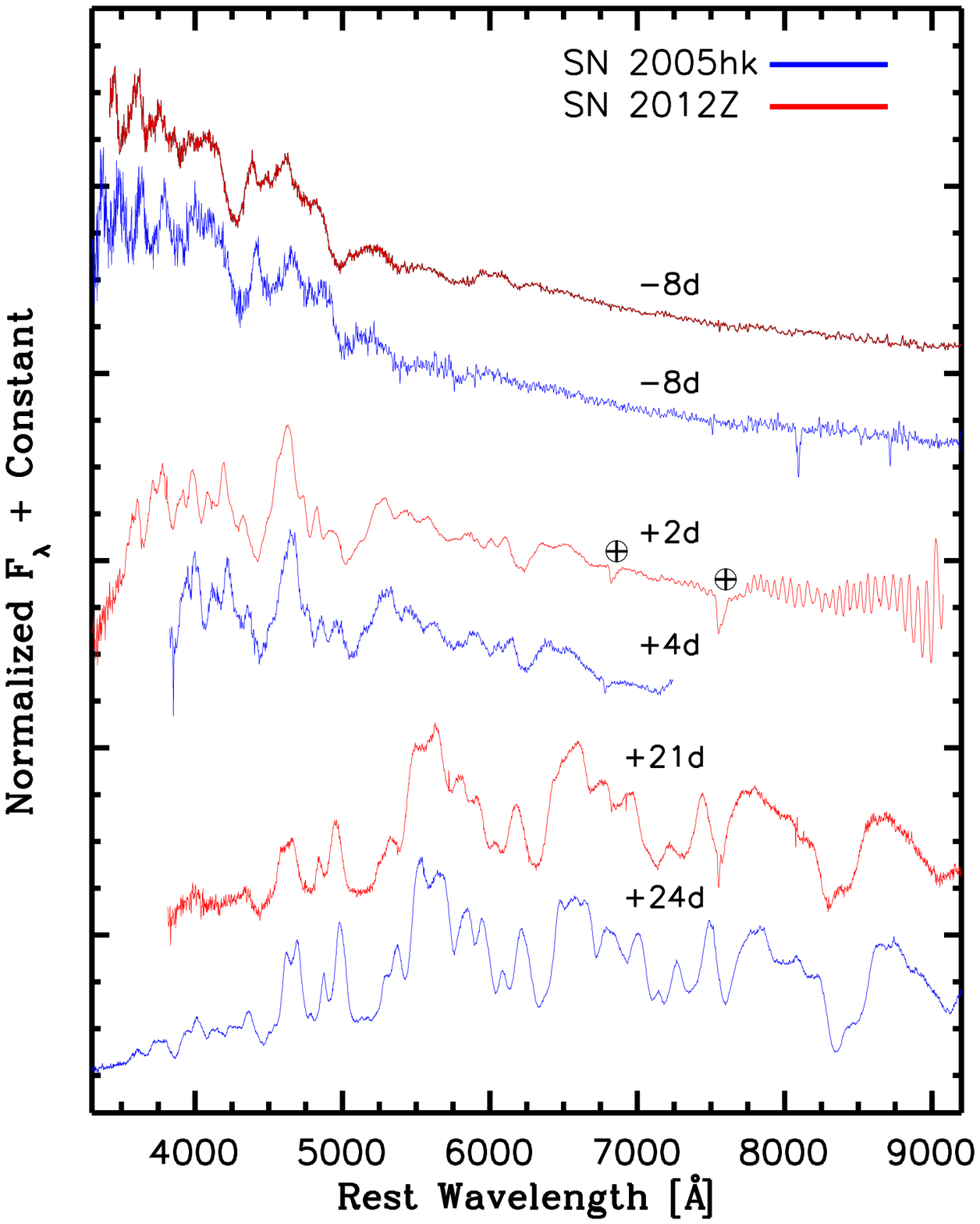}
\caption[]{Comparison of visual-wavelength spectra of SN~2012Z at phases 
of $-$8d, $+$2d, and $+$24d to similar epoch spectra of SN~2005hk  
\citep{phillips07}.
\label{optspeccomp}}
\end{figure}

\clearpage
\begin{figure}[h]
\centering
\includegraphics[width=5.6in]{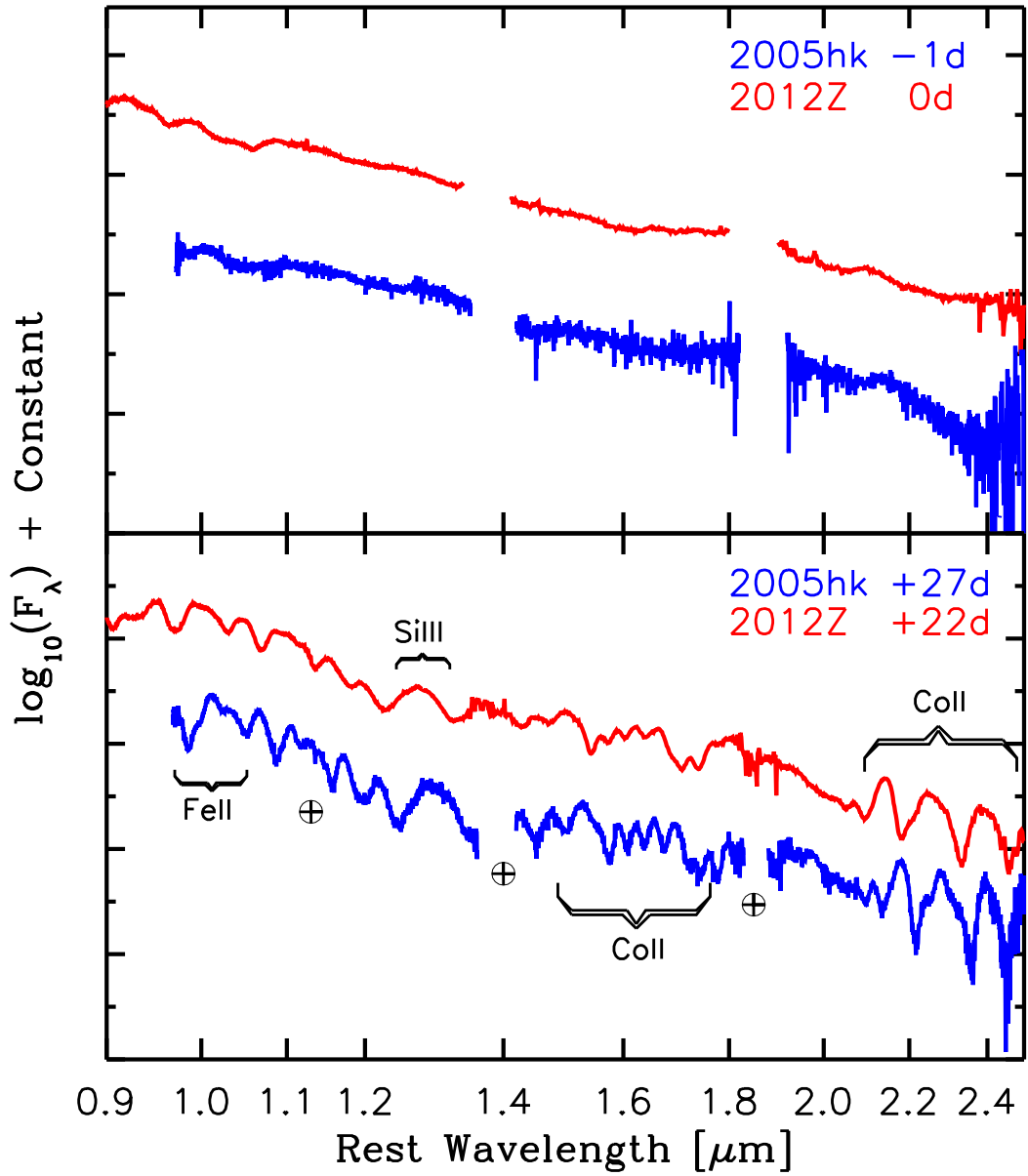}
\caption[]{Comparison of NIR-wavelength spectra of SN~2012Z at phases 
of 0d (top) and  $+$22d (bottom), to similar epoch spectra of SN~2005hk  
\citep{kromer13}. Prevalent features attributed to ions of
$\ion{Fe}{ii}$, $\ion{Si}{iii}$, and $\ion{Co}{ii}$ are indicated with
labels.  
 \label{nirspeccomp}}
\end{figure}

\clearpage
\begin{deluxetable}{crccc}
\tabletypesize{\normalsize}
\tablecolumns{5}
\tablenum{1}
\tablewidth{0pt}
\tablecaption{UVOT Ultraviolet Photometry of SN~2012Z.\label{table:uvot}}
\tablehead{
\colhead{JD--2450000}  &
\colhead{Phase$^{a}$}      &
\colhead{$uvw2$}   &
\colhead{$uvm2$}   &
\colhead{$uvw1$}   }
\startdata
5959.77    & $-$8.12 & 17.217(122) & 17.001(091) & 15.853(075) \\
5961.95    & $-$5.94 & 17.257(114) & 17.191(093) & 15.802(070) \\
5965.02    & $-$2.87 & 17.371(116) & 17.435(096) & 15.892(082) \\
5967.00    & $-$0.89 & 17.567(118) & 17.811(102) & 16.241(091) \\
5968.66    &    0.77 & 17.683(118) & 17.855(103) & 16.362(087) \\
5971.11    &    3.22 & $\cdots$    & $\cdots$    & 17.296(104) \\
5976.96    &    9.07 & $\cdots$    & 19.565(296) & $\cdots$     \\
5977.72    &    9.83 & 19.073(226) & $\cdots$    & 17.663(107) \\ 
5979.60    &   11.71 & 19.128(230) & $\cdots$    & 18.034(137) \\
5982.36    &   14.47 & 19.600(376) & $\cdots$    & 18.093(171) \\
5989.79    &   21.90 & $\cdots$    & $\cdots$    & 19.094(219)\\
5992.13    &   24.24 & $\cdots$    & $\cdots$    & 19.341(262)\\
5994.01    &   26.12 & $\cdots$    & $\cdots$    & 19.483(290)\\
5998.08    &   30.19 & $\cdots$    & $\cdots$    & 19.532(302)\\
\enddata
\tablecomments{One sigma uncertainties given in parentheses are in millimag.}
\tablenotetext{a}{Days past $T(B)_{max}$, i.e. JD--2455967.89.}
\end{deluxetable}

\clearpage
\begin{deluxetable}{ccccccccccccccccccccc}
\DTBLrotate
\tabletypesize{\scriptsize}
\tablecolumns{21}
\tablewidth{0pt}
\tablenum{2}
\tablecaption{Optical and NIR photometry of the local sequence of SN~2012Z in the {\it natural} system.\label{SN12Zlocalsequence}}
\tablehead{
\colhead{}     &
\colhead{}     &
\colhead{}     &
\colhead{$u$} &
\colhead{$N$} &
\colhead{$g$} &
\colhead{$N$} &
\colhead{$r$} &
\colhead{$N$} &
\colhead{$i$} &
\colhead{$N$} &
\colhead{$B$}  &
\colhead{$N$} &
\colhead{$V$}  &
\colhead{$N$} &
\colhead{$Y$}  &
\colhead{$N$} &
\colhead{$J$}  &
\colhead{$N$} &
\colhead{$H$} &
\colhead{$N$}  \\
\colhead{STAR}             &
\colhead{$\alpha~(2000)$}  &
\colhead{$\delta~(2000)$}  &
\colhead{(mag)}            &
\colhead{     }            &
\colhead{(mag)}            &
\colhead{     }            &
\colhead{(mag)}            &
\colhead{     }            &
\colhead{(mag)}            &
\colhead{     }            &
\colhead{(mag)}            &
\colhead{     }            &
\colhead{(mag)}            &
\colhead{     }            &
\colhead{(mag)}            &
\colhead{     }            &
\colhead{(mag)}            &
\colhead{     }            &
\colhead{(mag)}            &
\colhead{     }            }
\startdata
001  & 03:22:15.58 & -15:19:11.93 & 15.026(007) & 17       &  $\cdots$   & $\cdots$ &   $\cdots$   & $\cdots$ &   $\cdots$   & $\cdots$ &  13.339(005) &  7       &   $\cdots$ & $\cdots$   &  $\cdots$   & $\cdots$ &  $\cdots$   & $\cdots$  &  $\cdots$   & $\cdots$\\
002  & 03:22:19.50 & -15:19:19.63 & 14.038(006) & 18       &  $\cdots$   & $\cdots$ &   $\cdots$   & $\cdots$ &   $\cdots$   & $\cdots$ &  13.282(005) &  7       &  12.762(011) &  1       &  $\cdots$   & $\cdots$ &  $\cdots$   & $\cdots$  &  $\cdots$   & $\cdots$\\
003  & 03:22:00.97 & -15:19:28.24 & 14.358(006) & 18       &  $\cdots$   & $\cdots$ &   $\cdots$   & $\cdots$ &   $\cdots$   & $\cdots$ &  13.399(005) &  8       &  12.759(011) &  1       &  $\cdots$   & $\cdots$ &  $\cdots$   & $\cdots$  &  $\cdots$   & $\cdots$\\
004  & 03:21:52.18 & -15:20:52.44 & 17.045(010) & 19       & 16.002(003) & 22       &  15.636(003) & 20       &  15.501(003) & 25       &  16.304(003) & 26       &  15.791(003) & 28       &  $\cdots$   & $\cdots$ &  $\cdots$   & $\cdots$  &  $\cdots$   & $\cdots$\\
005  & 03:22:05.62 & -15:22:34.86 & 17.007(010) & 19       & 16.022(003) & 22       &  15.618(003) & 20       &  15.471(003) & 25       &  16.330(003) & 27       &  15.795(003) & 29       &  $\cdots$   & $\cdots$ &  $\cdots$   & $\cdots$  &  $\cdots$   & $\cdots$\\
006  & 03:21:54.34 & -15:21:12.06 & 17.220(011) & 19       & 16.101(003) & 22       &  15.648(003) & 20       &  15.477(003) & 25       &  16.435(003) & 27       &  15.849(003) & 29       &  $\cdots$   & $\cdots$ &  $\cdots$   & $\cdots$  &  $\cdots$   & $\cdots$\\
007  & 03:22:18.71 & -15:19:27.70 & 19.076(044) & 10       & 16.957(004) & 19       &  16.159(004) & 17       &  15.844(004) & 20       &  17.441(007) & 22       &  16.565(004) & 23       &  $\cdots$   & $\cdots$ &  $\cdots$   & $\cdots$  &  $\cdots$   & $\cdots$\\
008  & 03:22:09.62 & -15:22:13.87 & 18.311(020) & 18       & 17.188(004) & 22       &  16.780(005) & 20       &  16.616(005) & 25       &  17.532(006) & 27       &  16.965(004) & 29       &  $\cdots$   & $\cdots$ &  $\cdots$   & $\cdots$  &  $\cdots$   & $\cdots$\\
009  & 03:22:18.09 & -15:24:01.51 & $\cdots$    & $\cdots$ & 18.486(010) & 22       &  17.342(006) & 20       &  16.861(006) & 25       &  18.905(018) & 20       &  17.920(008) & 29       &  $\cdots$   & $\cdots$ &  $\cdots$   & $\cdots$  &  $\cdots$   & $\cdots$\\
010  & 03:21:53.30 & -15:27:14.29 & $\cdots$    & $\cdots$ & 19.155(024) & 11       &  17.840(012) & 12       &  16.165(006) & 15       &  19.778(071) &  5       &  18.510(018) & 17       &  $\cdots$   & $\cdots$ &  $\cdots$   & $\cdots$  &  $\cdots$   & $\cdots$\\
011  & 03:22:12.29 & -15:19:41.63 & $\cdots$    & $\cdots$ & 19.133(019) & 17       &  17.742(009) & 18       &  16.700(005) & 22       &  19.788(047) &  9       &  18.442(013) & 28       &  $\cdots$   & $\cdots$ &  $\cdots$   & $\cdots$  &  $\cdots$   & $\cdots$\\
012  & 03:22:16.89 & -15:20:44.88 & 18.917(082) &  1       & 18.880(014) & 21       &  18.060(011) & 20       &  17.628(010) & 25       &  18.962(024) & 14       &  18.463(014) & 29       &  $\cdots$   & $\cdots$ &  $\cdots$   & $\cdots$  &  $\cdots$   & $\cdots$\\
013  & 03:21:48.55 & -15:21:14.36 & 19.452(094) &  3       & 18.817(018) & 13       &  18.476(019) & 14       &  18.291(023) & 16       &  18.755(022) & 14       &  18.587(017) & 22       &  $\cdots$   & $\cdots$ &  $\cdots$   & $\cdots$  &  $\cdots$   & $\cdots$\\
014  & 03:21:52.45 & -15:27:02.48 & 18.958(048) &  7       & 18.991(017) & 16       &  18.666(019) & 17       &  18.674(031) & 18       &  18.970(025) & 13       &  18.881(023) & 20       &  $\cdots$   & $\cdots$ &  $\cdots$   & $\cdots$  &  $\cdots$   & $\cdots$\\
015  & 03:22:22.44 & -15:20:44.09 & $\cdots$    & $\cdots$ & 19.444(024) & 16       &  18.628(017) & 19       &  18.263(018) & 24       &  19.445(044) &  8       &  18.885(022) & 20       &  $\cdots$   & $\cdots$ &  $\cdots$   & $\cdots$  &  $\cdots$   & $\cdots$\\
016  & 03:21:55.41 & -15:20:24.11 & $\cdots$    & $\cdots$ & 20.111(044) & 12       &  18.848(021) & 19       &  17.568(010) & 25       &  20.147(104) &  3       &  19.480(036) & 15       &  $\cdots$   & $\cdots$ &  $\cdots$   & $\cdots$  &  $\cdots$   & $\cdots$\\
017  & 03:22:14.78 & -15:27:04.36 & $\cdots$    & $\cdots$ & 20.372(097) &  3       &  19.082(032) & 11       &  18.222(020) & 17       &   $\cdots$   & $\cdots$ &  19.656(059) &  7       &  $\cdots$   & $\cdots$ &  $\cdots$   & $\cdots$  &  $\cdots$   & $\cdots$\\
018  & 03:22:05.60 & -15:22:35.06 & $\cdots$    & $\cdots$ &  $\cdots$   & $\cdots$ &   $\cdots$   & $\cdots$ &   $\cdots$   & $\cdots$ &   $\cdots$   & $\cdots$ &   $\cdots$   & $\cdots$ & 14.808(016) &  4       & 14.572(017) &  5        & 14.246(014) &  5 \\
019  & 03:22:03.04 & -15:22:27.83 & $\cdots$    & $\cdots$ &  $\cdots$   & $\cdots$ &   $\cdots$   & $\cdots$ &   $\cdots$   & $\cdots$ &   $\cdots$   & $\cdots$ &   $\cdots$   & $\cdots$ & 16.431(024) &  4       & 16.035(028) &  5        & 15.384(023) &  5 \\
020  & 03:22:09.59 & -15:22:13.71 & $\cdots$    & $\cdots$ &  $\cdots$   & $\cdots$ &   $\cdots$   & $\cdots$ &   $\cdots$   & $\cdots$ &   $\cdots$   & $\cdots$ &   $\cdots$   & $\cdots$ & 15.973(020) &  4       & 15.708(024) &  5        & 15.398(022) &  5 \\
021  & 03:22:04.73 & -15:23:44.15 & $\cdots$    & $\cdots$ &  $\cdots$   & $\cdots$ &   $\cdots$   & $\cdots$ &   $\cdots$   & $\cdots$ &   $\cdots$   & $\cdots$ &   $\cdots$   & $\cdots$ & 16.357(033) &  4       & 15.901(032) &  5        & 15.245(029) &  5 \\
022  & 03:22:01.66 & -15:22:30.86 & $\cdots$    & $\cdots$ &  $\cdots$   & $\cdots$ &   $\cdots$   & $\cdots$ &   $\cdots$   & $\cdots$ &   $\cdots$   & $\cdots$ &   $\cdots$   & $\cdots$ & 18.399(096) &  2       &  $\cdots$   & $\cdots$  & $\cdots$    & $\cdots$\\
023  & 03:22:05.94 & -15:24:38.59 & $\cdots$    & $\cdots$ &  $\cdots$   & $\cdots$ &   $\cdots$   & $\cdots$ &   $\cdots$   & $\cdots$ &   $\cdots$   & $\cdots$ &   $\cdots$   & $\cdots$ & 17.229(100) &  1       & 16.854(104) &  1        & 16.383(119) &  1 \\
\enddata
\tablecomments{Uncertainties given in parentheses in thousandths of a
  magnitude correspond to an rms of the magnitudes obtained on
  photometric nights.}
\end{deluxetable}

\clearpage
\clearpage
\begin{deluxetable}{ccccccc}
\tablewidth{0pt}
\tabletypesize{\scriptsize}
\tablecaption{Swope optical photometry of SN~2012Z in the {\it natural} system.\label{SN12Zoptphot}}
\tablenum{3}
\tablehead{
\colhead{JD$-2,450,000+$} &
\colhead{$u$ (mag)}&
\colhead{$g$ (mag)}&
\colhead{$r$ (mag)}&
\colhead{$i$ (mag)}&
\colhead{$B$ (mag)} &
\colhead{$V$ (mag)} }
\startdata
5959.57 & 15.631(011) & 15.533(004) & 15.582(005) & 15.753(008) & 15.591(008) & 15.676(007) \\  
5960.57 & 15.448(008) & $\cdots$    & 15.322(009) & 15.488(009) & 15.368(009) & 15.416(009) \\
5961.57 & 15.323(006) & 15.106(004) & 15.124(005) & 15.282(005) & 15.171(008) & 15.214(008) \\
5962.56 & 15.233(007) & $\cdots$    & $\cdots$    & 15.115(005) & 15.027(006) & 15.028(005) \\
5963.58 & 15.190(011) & 14.839(004) & 14.814(004) & 14.985(004) & 14.925(007) & 14.896(004) \\
5964.60 & $\cdots$    & 14.746(006) & 14.704(006) & 14.871(008) & 14.857(006) & 14.785(005) \\
5965.61 & 15.165(016) & 14.697(006) & 14.608(008) & 14.783(008) & 14.812(009) & 14.698(007) \\
5966.59 & 15.156(013) & $\cdots$    & $\cdots$    & 14.698(006) & 14.759(006) & 14.619(005) \\
5967.59 & 15.199(012) & $\cdots$    & 14.472(006) & 14.623(007) & 14.750(007) & 14.562(005) \\
5968.59 & 15.254(008) & 14.584(007) & 14.417(006) & 14.578(008) & 14.753(004) & 14.512(006) \\
5969.59 & 15.321(009) & 14.572(004) & 14.369(004) & 14.517(008) & 14.780(004) & 14.479(007) \\
5970.56 & 15.425(009) & 14.579(004) & 14.336(005) & 14.467(007) & 14.830(008) & 14.458(008) \\
5971.57 & 15.520(010) & 14.586(007) & 14.288(009) & 14.431(009) & 14.829(006) & 14.412(006) \\
5972.57 & 15.629(009) & 14.646(004) & 14.287(005) & 14.423(008) & 14.915(004) & 14.422(006) \\
5973.55 & 15.751(009) & 14.698(004) & 14.279(004) & 14.389(005) & 15.007(005) & 14.428(006) \\
5974.57 & 15.893(007) & 14.750(004) & 14.268(004) & 14.370(004) & 15.086(004) & 14.437(004) \\
5979.58 & 16.859(009) & 15.235(005) & 14.322(005) & $\cdots$    & 15.699(007) & 14.671(006) \\
5985.54 & 17.998(015) & 15.934(004) & 14.604(004) & 14.470(004) & 16.498(006) & 15.121(005) \\
5986.57 & 18.138(017) & 16.063(008) & $\cdots$    & 14.517(004) & 16.601(007) & 15.204(005) \\
5988.57 & 18.350(024) & 16.211(005) & 14.800(004) & 14.608(004) & 16.792(007) & 15.364(005) \\
5990.58 & 18.496(051) & 16.352(006) & 14.925(005) & 14.716(005) & $\cdots$    & $\cdots$    \\
5991.58 & 18.672(076) & 16.422(008) & 14.983(005) & 14.769(004) & 17.018(011) & 15.559(004) \\
5992.56 & 18.656(041) & 16.511(006) & 15.047(008) & 14.811(006) & 17.074(011) & 15.643(008) \\
5993.53 & $\cdots$    & 16.562(006) & 15.100(004) & 14.875(004) & 17.129(009) & 15.694(005) \\
5994.54 & 18.764(067) & 16.599(007) & 15.157(004) & 14.922(006) & 17.168(011) & 15.749(006) \\
5995.54 & 18.880(064) & 16.668(008) & 15.215(006) & 14.969(007) & 17.237(010) & 15.797(007) \\
5996.53 & 18.945(038) & 16.713(009) & 15.267(008) & 15.023(008) & 17.250(014) & 15.848(006) \\
5997.52 & 19.017(023) & 16.743(007) & 15.310(007) & 15.072(008) & 17.299(009) & 15.890(006) \\
5998.52 & 19.063(030) & 16.773(004) & 15.349(004) & 15.101(004) & 17.347(009) & 15.923(006) \\
5999.52 & 19.024(021) & 16.819(011) & 15.396(009) & 15.168(010) & 17.375(010) & 15.994(009) \\
6000.56 & $\cdots$    & 16.826(008) & 15.418(004) & 15.172(006) & 17.395(008) & 15.995(004) \\
6001.55 & $\cdots$    & 16.847(005) & 15.467(004) & 15.214(005) & 17.419(009) & 16.024(007) \\
6002.55 & $\cdots$    & 16.853(005) & 15.502(004) & 15.252(006) & 17.422(009) & 16.045(006) \\
6003.55 & $\cdots$    & 16.877(005) & 15.530(004) & 15.275(004) & 17.451(009) & 16.085(004) \\
6004.55 & $\cdots$    & 16.912(006) & 15.563(004) & 15.316(004) & 17.474(009) & 16.103(004) \\
6005.55 & $\cdots$    & $\cdots$    & $\cdots$    & 15.341(004) & 17.478(009) & 16.121(005) \\
6006.54 & $\cdots$    & 16.954(008) & 15.623(006) & 15.385(008) & 17.520(009) & 16.157(006) \\
6007.52 & $\cdots$    & 16.975(009) & 15.676(009) & 15.404(009) & 17.522(009) & 16.188(009) \\
6009.52 & $\cdots$    & 16.978(006) & 15.702(006) & 15.452(006) & 17.545(007) & 16.213(006) \\
6010.51 & $\cdots$    & 17.000(007) & 15.741(006) & 15.498(008) & 17.551(008) & 16.235(007) \\
6011.52 & $\cdots$    & $\cdots$    & $\cdots$    & $\cdots$    & 17.596(010) & 16.253(007) \\
6012.53 & $\cdots$    & $\cdots$    & $\cdots$    & $\cdots$    & 17.581(009) & 16.273(005) \\
6014.51 & $\cdots$    & 17.056(007) & 15.850(005) & 15.606(005) & $\cdots$    & $\cdots$    \\
6015.53 & $\cdots$    & $\cdots$    & $\cdots$    & $\cdots$    & $\cdots$    & 16.335(008) \\
6017.52 & $\cdots$    & $\cdots$    & $\cdots$    & $\cdots$    & 17.625(015) & 16.358(007) \\
6018.49 & $\cdots$    & 17.097(010) & 15.942(004) & 15.714(004) & $\cdots$    & $\cdots$    \\
6019.49 & $\cdots$    & $\cdots$    & $\cdots$    & $\cdots$    & 17.644(014) & 16.386(008) \\
6233.82 & $\cdots$    & $\cdots$    & $\cdots$    & 20.381(119) & 21.411(318) & 20.876(206) \\
\enddata
\tablecomments{Values in parentheses are 1$\sigma$ measurement uncertainties in millimag.}
\end{deluxetable}

\clearpage
\begin{deluxetable}{ccccc}
\tablewidth{0pt}
\tablenum{4}
\tablecaption{du Pont NIR photometry of SN~2012Z in the {\it natural} system.\label{SN12Znirphot}}
\tablehead{
\colhead{JD$-$$2,450,000+$} &
\colhead{$Y$ (mag)}&
\colhead{$J$ (mag)}&
\colhead{$H$ (mag)}}
\startdata
5992.51541 & 13.899(013)& 14.208(013) & 13.963(012) \\   
5996.52808 & $\cdots$   & $\cdots$    & 14.119(011) \\ 
5997.49164 & 14.047(012)& 14.402(013) & $\cdots$    \\
6000.49606 & 14.139(012)& 14.557(013) & 14.288(011) \\
6001.49526 & 14.185(013)& 14.565(013) & 14.277(011) \\
6022.46819 & 14.912(011)& $\cdots$    & $\cdots$    \\
6167.85063 & 19.818(055)& 19.269(046) & $\cdots$    \\
6168.82954 & $\cdots$   & 19.160(065) & $\cdots$    \\
\enddata
\tablecomments{Values in parentheses are 1$\sigma$ measurement uncertainties in millimag.}
\end{deluxetable}

 \clearpage
 \begin{deluxetable}{ccccc}
\tabletypesize{\normalsize}
\tablewidth{0pt}
\tablenum{5}
\tablecaption{Journal of Spectroscopic Observations.\label{specjor}}
\tablehead{
\colhead{Date} &
\colhead{JD$-$2,450,000+} &
\colhead{Phase$^{a}$} &
\colhead{Telescope} &
\colhead{Instrument}} 
\startdata
\multicolumn{5}{c}{\bf Optical}\\
2012 Feb 01.2 & 5958.7 & $-$9.2   & Lick   & Kast                   \\ 
2012 Feb 02.2 & 5959.7 & $-$8.2   & Lick   & Kast                   \\
2012 Feb 04.1 & 5961.5 & $-$6.4   & Clay   & MIKE                   \\
2012 Feb 12.2 & 5969.7 & $+$1.8   & NOT    & Alfosc                 \\
2012 Feb 16.1 & 5973.6 & $+$5.7   & FLWO   & FAST                   \\
2012 Feb 18.9 & 5976.4 & $+$8.5   & NOT    & Alfosc                 \\
2012 Feb 19.1 & 5976.6 & $+$8.7   & FLWO   & FAST                   \\
2012 Feb 21.0 & 5978.6 & $+$10.7  & FLWO   & FAST                   \\
2012 Feb 21.1 & 5979.6 & $+$11.7  & FLWO   & FAST                   \\
2012 Feb 22.1 & 5980.5 & $+$12.6  & Clay   & LDSS3                  \\
2012 Feb 25.8 & 5983.4 & $+$15.5  & NOT    & Alfosc                 \\
2012 Mar 02.1 & 5988.6 & $+$20.7  & Clay   & LDSS3                  \\
2012 Mar 15.2 & 6001.7 & $+$33.8  & Keck   & LRIS                   \\
2012 Aug 22.1 & 6161.6 & $+$193.7 & SALT   & RSS                    \\
2012 Sep 12.3 & 6182.8 & $+$214.9 & du Pont& WFCCD                  \\
2012 Oct 15.5 & 6216.0 & $+$248.1 & Keck   & DEIMOS                 \\
\multicolumn{5}{c}{\bf NIR}\\
2012 Feb 03.0 & 5960.6 & $-$7.3   & Baade  & Fire                   \\
2012 Feb 10.0 & 5967.5 & $-$0.4   & Baade  & Fire                   \\
2012 Feb 15.0 & 5972.5 & $+$4.6   & NTT    & Sofi                   \\
2012 Feb 21.0 & 5978.5 & $+$10.6  & VLT    & ISAAC                  \\ 
2012 Feb 27.0 & 5984.5 & $+$16.6  & VLT    & ISAAC                  \\
2012 Mar 03.0 & 5989.5 & $+$21.6  & Baade  & Fire                   \\
2012 Mar 14.9 & 6001.5 & $+$33.6  & VLT    & ISAAC                  \\ 
2012 Mar 29.9 & 6016.5 & $+$48.6  & VLT    & ISAAC                  \\
2012 Apr 07.9 & 6025.5 & $+$57.6  & Baade  & Fire                   \\
2012 Nov 05.2 & 6236.7 & $+$268.8 & Baade  & Fire                   \\ 
\enddata
\tablenotetext{a}{Days since $T(B)_{max}$, i.e., JD$-$2455967.8.}
\end{deluxetable}

\clearpage
\begin{deluxetable} {lcclc}
\tablecolumns{5}
\tablenum{6}
\tablewidth{0pc}
\tablecaption{Optical Light Curve Parameters of SNe~2005hk and 2012Z.\label{lcpar}}
\tablehead{
\colhead{Filter} &
\colhead{Peak Time} &
\colhead{Peak Obs.} &
\colhead{$\Delta$$m_{15}$} &
\colhead{Peak Abs.} \\ 
\colhead{} &
\colhead{(JD$-$2,450,000)} &
\colhead{(mag)} &
\colhead{(mag)} &
\colhead{(mag)} }
\startdata
\multicolumn{5}{c}{\bf SN 2005hk}\\
$u$ & 3682.54$\pm$0.36 & 16.27$\pm$0.01 & 1.96$\pm$0.07 & $-$17.73$\pm$0.25\\
$g$ & 3685.92$\pm$0.08 & 15.78$\pm$0.01 & 1.36$\pm$0.01 & $-$18.08$\pm$ 0.25\\
$r$ & 3691.66$\pm$0.23 & 15.68$\pm$0.01 & 0.70$\pm$0.02 & $-$18.07$\pm$0.25   \\
$i$ & 3695.27$\pm$0.16 & 15.80$\pm$0.01 & 0.60$\pm$0.01 & $-$17.88$\pm$0.25   \\
$B$ & 3685.16$\pm$0.07 & 15.91$\pm$0.01 & 1.62$\pm$0.01 & $-$18.00$\pm$0.25   \\
$V$ & 3688.28$\pm$0.14 & 15.73$\pm$0.01 & 0.92$\pm$0.01 & $-$18.07$\pm$0.25   \\
\multicolumn{5}{c}{\bf SN 2012Z}\\
$u$ & 5965.48$\pm$0.35 & 15.16$\pm$0.01 & 1.92$\pm$0.08 & $-$17.95$\pm$0.09   \\          
$g$ & 5969.78$\pm$0.12 & 14.57$\pm$0.01 & 1.30$\pm$0.01 & $-$18.40$\pm$0.09   \\
$r$ & 5975.26$\pm$0.33 & 14.27$\pm$0.01 & 0.66$\pm$0.02 & $-$18.60$\pm$0.09   \\
$i$ & 5978.46$\pm$0.53 & 14.34$\pm$0.01 & 0.54$\pm$0.04 & $-$18.46$\pm$0.09   \\  
$B$ & 5967.89$\pm$0.11 & 14.75$\pm$0.01 & 1.43$\pm$0.02 & $-$18.27$\pm$0.09   \\
$V$ & 5972.74$\pm$0.14 & 14.42$\pm$0.01 & 0.89$\pm$0.01 & $-$18.50$\pm$0.09   \\
\enddata
\end{deluxetable}

\clearpage
\appendix
\section{Definitive Photometry of SN~2005hk}

A portion of the CSP photometry of SN~2005hk originally published by
\citet{phillips07} was computed from science images which underwent
template subtraction using provisional template images.  We returned
to the field of UGC~272 and obtained updated optical $BV$- and NIR
$YJHK_s$-band images. These were used to compute proper template
subtraction for the respective science images.  In addition, the field
of local sequence stars was also re-calibrated based on a larger
number of nights in which standard star fields were observed.
Table~\ref{SN05hklocalsequence} contains the re-calibration of our
optical and NIR local sequences, while our revised optical and NIR
photometry of SN~2005hk is presented in Table~\ref{SN05hkoptphot} and
Table~\ref{SN05hknirphot}, respectively. Note that all of our
photometry of the local sequence stars and the SN are presented in the
{\it natural} system \citep{stritzinger11}.

\clearpage
\begin{deluxetable}{ccccccccccccccccccccccc}
\DTBLrotate
\tabletypesize{\scriptsize}
\tablecolumns{23}
\tablewidth{0pt}
\tablenum{A1}
\tablecaption{Optical and NIR photometry of the local sequence of SN~2005hk in the {\it natural} system.\label{SN05hklocalsequence}}
\tablehead{
\colhead{}     &
\colhead{}     &
\colhead{}     &
\colhead{$u$} &
\colhead{$N$} &
\colhead{$g$} &
\colhead{$N$} &
\colhead{$r$} &
\colhead{$N$} &
\colhead{$i$} &
\colhead{$N$} &
\colhead{$B$} &
\colhead{$N$} &
\colhead{$V$} &
\colhead{$N$} &
\colhead{$Y$} &
\colhead{$N$} &
\colhead{$J$} &
\colhead{$N$} &
\colhead{$H$} &
\colhead{$N$} &
\colhead{$K_s$} &
\colhead{$N$}  \\
\colhead{STAR}             &
\colhead{$\alpha~(2000)$}  &
\colhead{$\delta~(2000)$}  &
\colhead{(mag)}            &
\colhead{     }            &
\colhead{(mag)}            &
\colhead{     }            &
\colhead{(mag)}            &
\colhead{     }            &
\colhead{(mag)}            &
\colhead{     }            &
\colhead{(mag)}            &
\colhead{     }            &
\colhead{(mag)}            &
\colhead{     }            &
\colhead{(mag)}            &
\colhead{     }            &
\colhead{(mag)}            &
\colhead{     }            &
\colhead{(mag)}            &
\colhead{     }            &
\colhead{(mag)}            &
\colhead{     }            }
\startdata
001 & 00:27:58.40 & -01:13:12.19 & 15.556(019) & 5        & 14.458(009) & 2        & 14.499(007) & 3         & 14.590(006) & 4         & 14.602(006) & 5         & 14.441(006) & 4         & 12.692(008) & 5        & 12.480(007) & 5        & 12.157(007) & 5        & $\cdots$    & $\cdots$\\
002 & 00:27:59.34 & -01:09:36.38 & 16.625(023) & 5        & 15.020(007) & 3        & 14.420(009) & 2         & 14.201(010) & 1         & 15.422(006) & 5         & 14.673(006) & 4         & 13.467(009) & 4        & 13.226(008) & 5        & 12.848(008) & 5        & $\cdots$    & $\cdots$\\  
003 & 00:28:01.15 & -01:11:29.09 & 17.883(037) & 4        & 16.459(006) & 5        & 15.820(006) & 5         & 15.582(006) & 5         & 16.858(009) & 5         & 16.090(008) & 5         & 14.824(010) & 5        & 14.530(008) & 5        & 14.114(009) & 5        & $\cdots$    & $\cdots$\\
004 & 00:28:02.93 & -01:11:23.56 & 19.259(164) & 1        & 17.086(007) & 5        & 15.859(006) & 5         & 15.355(006) & 5         & 17.663(014) & 5         & 16.462(009) & 5         & 14.344(009) & 5        & 13.970(007) & 5        & 13.394(007) & 5        & $\cdots$    & $\cdots$\\
005 & 00:27:38.83 & -01:09:16.30 & 17.641(036) & 4        & 16.521(006) & 5        & 15.953(007) & 5         & 15.733(006) & 5         & 16.879(009) & 5         & 16.199(009) & 5         & $\cdots$    & $\cdots$ & $\cdots$    & $\cdots$ & $\cdots$    & $\cdots$ & $\cdots$    & $\cdots$\\
006 & 00:27:38.38 & -01:11:30.82 & 18.537(060) & 3        & 16.723(006) & 5        & 15.911(006) & 5         & 15.575(006) & 5         & 17.180(011) & 5         & 16.283(009) & 5         & $\cdots$    & $\cdots$ & $\cdots$    & $\cdots$ & $\cdots$    & $\cdots$ & $\cdots$    & $\cdots$\\
007 & 00:27:51.58 & -01:14:36.05 & 17.245(027) & 5        & 16.258(006) & 5        & 16.057(007) & 5         & 16.036(007) & 5         & 16.471(008) & 5         & 16.127(008) & 5         & 15.497(016) & 4        & 15.361(013) & 5        & 15.209(019) & 5        & $\cdots$    & $\cdots$\\
008 & 00:27:50.88 & -01:08:02.40 & 17.669(032) & 4        & 16.706(006) & 5        & 16.334(007) & 5         & 16.165(007) & 5         & 17.009(010) & 5         & 16.467(009) & 5         & 15.543(017) & 4        & 15.347(017) & 5        & 15.007(022) & 5        & $\cdots$    & $\cdots$\\
009 & 00:28:00.54 & -01:15:50.50 &  $\cdots$   & $\cdots$ & 17.609(009) & 4        & 16.492(008) & 4         & 16.052(008) & 4         & 18.175(019) & 4         & 17.058(012) & 4         & 15.092(015) & 4        & 14.722(012) & 5        & 14.123(013) & 5        & $\cdots$    & $\cdots$\\
010 & 00:27:47.46 & -01:07:55.83 & 18.483(057) & 3        & 17.288(007) & 5        & 16.685(008) & 5         & 16.452(009) & 5         & 17.696(014) & 5         & 16.949(011) & 5         & 15.651(018) & 4        & 15.417(017) & 5        & 15.060(023) & 5        & $\cdots$    & $\cdots$\\
011 & 00:27:47.82 & -01:13:59.88 & 19.505(198) & 1        & 17.683(009) & 5        & 16.823(008) & 5         & 16.469(009) & 5         & 18.109(018) & 5         & 17.238(012) & 5         & 15.624(013) & 5        & 15.357(010) & 6        & 14.836(014) & 6        & 14.652(056) & 1 \\
012 & 00:28:02.72 & -01:08:07.84 & 19.271(116) & 2        & 17.845(010) & 5        & 17.009(009) & 5         & 16.686(010) & 5         & 18.332(022) & 5         & 17.448(014) & 5         & 15.821(020) & 4        & 15.553(019) & 5        & 14.982(022) & 5        & $\cdots$    & $\cdots$\\
013 & 00:27:58.56 & -01:12:13.40 & 18.957(082) & 3        & 17.640(009) & 5        & 17.019(009) & 5         & 16.737(010) & 5         & 18.031(017) & 5         & 17.298(013) & 5         & 15.910(015) & 5        & 15.656(014) & 5        & 15.212(019) & 5        & $\cdots$    & $\cdots$\\ 
014 & 00:28:05.55 & -01:14:45.17 & 18.346(049) & 3        & 17.413(008) & 5        & 17.031(009) & 5         & 16.895(011) & 5         & 17.727(014) & 5         & 17.194(012) & 5         & 16.244(024) & 4        & 16.005(023) & 5        & 15.787(035) & 5        & $\cdots$    & $\cdots$\\
015 & 00:27:41.73 & -01:10:57.01 & 19.363(187) & 1        & 17.871(010) & 5        & 17.221(010) & 5         & 16.951(012) & 5         & 18.263(021) & 5         & 17.511(014) & 5         & $\cdots$    & 0        & $\cdots$    & $\cdots$ & $\cdots$    & $\cdots$ & $\cdots$    & $\cdots$\\
016 & 00:28:05.45 & -01:12:24.64 &  $\cdots$   & $\cdots$ & 18.848(022) & 4        & 17.466(012) & 5         & 16.602(010) & 5         & 19.437(055) & 3         & 18.109(022) & 5         & 15.400(012) & 5        & 14.998(010) & 5        & 14.408(011) & 5        & $\cdots$    & $\cdots$\\
017 & 00:27:47.99 & -01:12:55.62 &  $\cdots$   & $\cdots$ & 18.556(016) & 5        & 17.719(014) & 5         & 17.356(015) & 5         & 19.006(037) & 3         & 18.116(020) & 5         & 16.467(023) & 5        & 16.170(017) & 6        & 15.545(023) & 6        & 16.200(163) & 1 \\
018 & 00:27:38.35 & -01:09:00.28 &  $\cdots$   & $\cdots$ & 18.886(022) & 4        & 17.588(013) & 5         & 16.130(007) & 5         & 19.637(066) & 3         & 18.254(023) & 5         & $\cdots$    & 0        & $\cdots$    & $\cdots$ & $\cdots$    & $\cdots$ & $\cdots$    & $\cdots$\\
019 & 00:27:37.32 & -01:10:55.99 &  $\cdots$   & $\cdots$ & 19.093(026) & 4        & 17.845(015) & 5         & 17.237(014) & 5         & 19.672(067) & 3         & 18.313(024) & 5         & $\cdots$    & 0        & $\cdots$    & $\cdots$ & $\cdots$    & $\cdots$ & $\cdots$    & $\cdots$\\
020 & 00:27:55.68 & -01:14:38.15 & 18.003(053) & 3        & 18.467(015) & 5        & 18.104(018) & 5         & 17.856(024) & 5         & 18.714(029) & 3         & 18.279(024) & 5         & $\cdots$    & $\cdots$ & $\cdots$    & $\cdots$ & $\cdots$    & $\cdots$ & $\cdots$    & $\cdots$\\
021 & 00:27:50.49 & -01:09:04.46 & 18.805(107) & 1        & 18.844(021) & 4        & 18.504(025) & 5         & 18.314(035) & 5         & 19.093(041) & 4         & 18.658(032) & 5         & $\cdots$    & $\cdots$ & $\cdots$    & $\cdots$ & $\cdots$    & $\cdots$ & $\cdots$    & $\cdots$\\
022 & 00:27:55.61 & -01:07:57.62 &  $\cdots$   & $\cdots$ & 19.220(029) & 4        & 18.886(034) & 5         & 18.649(050) & 4         & 19.546(060) & 3         & 19.099(045) & 3         & $\cdots$    & $\cdots$ & $\cdots$    & $\cdots$ & $\cdots$    & $\cdots$ & $\cdots$    & $\cdots$\\
023 & 00:27:36.28 & -01:11:58.38 &  $\cdots$   & $\cdots$ & 19.911(054) & 3        & 18.732(029) & 4         & 17.859(025) & 4         & 20.634(198) & 1         & 19.533(072) & 3         & $\cdots$    & $\cdots$ & $\cdots$    & $\cdots$ & $\cdots$    & $\cdots$ & $\cdots$    & $\cdots$\\
024 & 00:27:58.91 & -01:15:00.70 &  $\cdots$   & $\cdots$ & 20.097(064) & 3        & 18.748(031) & 4         & 17.793(022) & 5         & $\cdots$    & $\cdots$  & 18.482(041) & 4         & 16.614(032) & 4        & 16.169(027) & 5        & 15.667(033) & 5        & $\cdots$    & $\cdots$\\
025 & 00:27:46.17 & -01:09:55.61 &  $\cdots$   & $\cdots$ & 20.380(093) & 2        & 19.105(042) & 4         & 18.224(032) & 5         & $\cdots$    & $\cdots$  & 19.740(084) & 3         & $\cdots$    & $\cdots$ & $\cdots$    & $\cdots$ & $\cdots$    & $\cdots$ & $\cdots$    & $\cdots$\\
026 & 00:27:57.20 & -01:13:03.18 &  $\cdots$   & $\cdots$ & $\cdots$    & $\cdots$ &  $\cdots$   & $\cdots$  &  $\cdots$   & $\cdots$  & $\cdots$    & $\cdots$  & $\cdots$    & $\cdots$  & 12.692(008) & 5        & 12.480(007) & 5        & 12.157(007) & 5        & $\cdots$    & $\cdots$\\
027 & 00:28:04.74 & -01:12:35.09 &  $\cdots$   & $\cdots$ & $\cdots$    & $\cdots$ &  $\cdots$   & $\cdots$  &  $\cdots$   & $\cdots$  & $\cdots$    & $\cdots$  & $\cdots$    & $\cdots$  & 13.077(008) & 5        & 12.848(007) & 5        & 12.498(007) & 5        & $\cdots$    & $\cdots$\\
028 & 00:28:02.42 & -01:09:21.58 &  $\cdots$   & $\cdots$ & $\cdots$    & $\cdots$ &  $\cdots$   & $\cdots$  &  $\cdots$   & $\cdots$  & $\cdots$    & $\cdots$  & $\cdots$    & $\cdots$  & 13.261(010) & 4        & 12.975(008) & 5        & 12.510(009) & 5        & $\cdots$    & $\cdots$\\
029 & 00:27:47.38 & -01:13:05.76 &  $\cdots$   & $\cdots$ & $\cdots$    & $\cdots$ &  $\cdots$   & $\cdots$  &  $\cdots$   & $\cdots$  & $\cdots$    & $\cdots$  & $\cdots$    & $\cdots$  & 16.482(023) & 5        & 16.063(016) & 6        & 15.492(022) & 6        & 15.631(116) & 1\\
030 & 00:28:06.38 & -01:15:11.20 &  $\cdots$   & $\cdots$ & $\cdots$    & $\cdots$ &  $\cdots$   & $\cdots$  &  $\cdots$   & $\cdots$  & $\cdots$    & $\cdots$  & $\cdots$    & $\cdots$  & 16.239(024) & 4        & 15.852(021) & 5        & 15.343(026) & 5        & $\cdots$    & $\cdots$\\
031 & 00:28:02.09 & -01:08:55.42 &  $\cdots$   & $\cdots$ & $\cdots$    & $\cdots$ &  $\cdots$   & $\cdots$  &  $\cdots$   & $\cdots$  & $\cdots$    & $\cdots$  & $\cdots$    & $\cdots$  & 16.850(042) & 4        & 16.548(043) & 5        & 16.058(057) & 5        & $\cdots$    & $\cdots$\\
032 & 00:28:12.89 & -01:08:33.94 &  $\cdots$   & $\cdots$ & $\cdots$    & $\cdots$ &  $\cdots$   & $\cdots$  &  $\cdots$   & $\cdots$  & $\cdots$    & $\cdots$  & $\cdots$    & $\cdots$  & 15.474(017) & 4        & 15.056(014) & 5        & 14.456(016) & 5        & $\cdots$    & $\cdots$\\
033 & 00:27:45.47 & -01:10:09.93 &  $\cdots$   & $\cdots$ & $\cdots$    & $\cdots$ &  $\cdots$   & $\cdots$  &  $\cdots$   & $\cdots$  & $\cdots$    & $\cdots$  & $\cdots$    & $\cdots$  & 16.803(036) & 4        & 16.356(028) & 4        & 15.893(049) & 3        & $\cdots$    & $\cdots$\\
034 & 00:28:06.44 & -01:07:07.95 &  $\cdots$   & $\cdots$ & $\cdots$    & $\cdots$ &  $\cdots$   & $\cdots$  &  $\cdots$   & $\cdots$  & $\cdots$    & $\cdots$  & $\cdots$    & $\cdots$  & 15.414(020) & 3        & 15.050(016) & 4        & 14.467(017) & 4        & $\cdots$    & $\cdots$\\
035 & 00:28:13.85 & -01:09:28.74 &  $\cdots$   & $\cdots$ & $\cdots$    & $\cdots$ &  $\cdots$   & $\cdots$  &  $\cdots$   & $\cdots$  & $\cdots$    & $\cdots$  & $\cdots$    & $\cdots$  & 15.082(015) & 3        & 14.704(011) & 5        & 14.120(012) & 5        & $\cdots$    & $\cdots$\\
036 & 00:27:51.47 & -01:13:23.41 &  $\cdots$   & $\cdots$ & $\cdots$    & $\cdots$ &  $\cdots$   & $\cdots$  &  $\cdots$   & $\cdots$  & $\cdots$    & $\cdots$  & $\cdots$    & $\cdots$  & 17.227(043) & 5        & 16.855(031) & 6        & 16.338(046) & 5        & $\cdots$    & $\cdots$\\
\enddata
\tablecomments{Uncertainties given in parentheses in thousandths of a
  magnitude correspond to an rms of the magnitudes obtained on
  photometric nights.}
\end{deluxetable}

\clearpage
\clearpage
\begin{deluxetable}{ccccccc}
\tablewidth{0pt}
\tabletypesize{\scriptsize}
\tablecaption{Revised CSP optical photometry of SN~2005hk in the {\it natural} system.\label{SN05hkoptphot}}
\tablenum{A2}
\tablehead{
\colhead{JD$-2,453,000+$} &
\colhead{$u$ (mag)}&
\colhead{$g$ (mag)}&
\colhead{$r$ (mag)}&
\colhead{$i$ (mag)}&
\colhead{$B$ (mag)} &
\colhead{$V$ (mag)} }
\startdata
675.61 & 16.759(016)  & 16.759(008) & 16.893(009) & 17.106(012) & 16.830(009) & 16.857(010) \\ 
682.58 & 16.265(014)  & 15.883(004) & 15.945(005) & 16.170(005) & 15.978(006) & 15.933(006) \\
683.58 & 16.280(017)  & 15.828(006) & 15.878(007) & 16.104(009) & 15.930(006) & 15.871(007) \\
684.62 & 16.326(018)  & 15.801(008) & 15.833(009) & 16.064(010) & 15.920(007) & 15.821(007) \\
687.60 & 16.541(019)  & 15.804(008) & 15.734(008) & 15.943(010) & 15.967(000) & 15.727(009) \\
689.66 & 16.774(024)  & 15.886(009) & 15.701(007) & 15.885(009) & 16.118(009) & 15.761(008) \\
690.61 & 16.917(023)  & 15.958(006) & 15.684(004) & 15.870(005) & 16.185(008) & 15.759(007) \\
694.65 & 17.619(026)  & 16.327(006) & 15.709(006) & 15.779(009) & 16.688(009) & 15.928(008) \\
695.64 & 17.812(029)  & 16.446(007) & 15.726(008) & 15.783(008) & 16.829(008) & 15.976(007) \\
698.59 & 18.349(041)  & 16.827(005) & 15.869(004) & 15.846(005) & 17.286(009) & 16.235(006) \\
699.62 & 18.688(056)  & 16.964(005) & 15.924(004) & 15.875(006) & 17.428(009) & 16.319(007) \\
702.63 & 19.083(066)  & 17.310(007) & 16.108(006) & 16.005(008) & 17.793(014) & 16.580(008) \\
706.66 & 19.538(067)  & 17.650(008) & 16.376(009) & 16.210(009) & 18.166(015) & 16.884(009) \\
712.57 & 19.896(123)  & 17.961(012) & 16.669(009) & 16.495(010) & 18.429(018) & 17.177(011) \\
720.57 & 20.265(288)  & 18.134(015) & 16.966(007) & 16.811(009) & 18.647(030) & 17.468(015) \\
725.60 & 20.261(117)  & 18.272(011) & 17.127(008) & 16.982(010) & 18.766(020) & 17.569(012) \\
728.61 & $\cdots$     & 18.332(009) & 17.202(008) & 17.066(009) & 18.823(022) & 17.663(010) \\
730.62 & 20.394(121)  & 18.309(009) & 17.267(007) & 17.135(008) & 18.797(021) & 17.697(011) \\
736.57 & 20.508(071)  & 18.411(007) & 17.396(007) & 17.289(008) & 18.875(015) & 17.782(009) \\
739.57 & 20.373(085)  & 18.466(008) & 17.475(007) & 17.367(009) & 18.933(015) & 17.839(010) \\
745.54 & $\cdots$     & 18.498(016) & 17.588(011) & 17.500(012) & 18.951(048) & 17.928(019) \\
748.56 & $\cdots$     & 18.528(016) & 17.657(012) & 17.564(013) & 19.034(067) & 17.993(024) \\
751.55 & $\cdots$     & 18.572(011) & 17.733(011) & 17.666(014) & 19.069(028) & 18.031(017) \\
761.53 & $\cdots$     & 18.665(019) & 17.983(010) & 17.841(012) & 19.194(026) & 18.181(013) \\
\enddata
\tablecomments{Values in parentheses are 1$\sigma$ measurement uncertainties in millimag.}
\end{deluxetable}

\clearpage
\begin{deluxetable}{ccccc}
\tablewidth{0pt}
\tablenum{A3}
\tablecaption{Revised NIR photometry of SN~2005hk in the {\it natural} system.\label{SN05hknirphot}}
\tablehead{
\colhead{JD$-2,453,000+$} &
\colhead{$Y$ (mag)}&
\colhead{$J$ (mag)}&
\colhead{$H$ (mag)} &
\colhead{$K_s$ (mag)}}
\startdata
676.49 & 16.818(024) & 16.780(031) & 16.929(061) & $\cdots$ \\ 
677.52 & 16.660(021) & 16.545(028) & 16.759(050) & $\cdots$ \\
678.51 & 16.498(019) & 16.434(027) & 16.593(064) & $\cdots$ \\
680.52 & 16.294(016) & 16.177(024) & 16.390(058) & $\cdots$ \\
681.57 & 16.225(013) & 16.167(012) & 16.280(016) & 16.208(054) \\  
685.60 & 16.024(021) & 15.963(021) & 16.083(038) & $\cdots$ \\
688.63 & $\cdots$    & 15.912(020) & 15.907(025) & $\cdots$ \\
691.60 & $\cdots$    & 15.908(056) & $\cdots$    & $\cdots$ \\
692.58 & $\cdots$    & $\cdots$    & 15.776(026) & $\cdots$ \\
696.61 & 15.531(016) & 15.888(020) & 15.654(021) & $\cdots$ \\
697.55 & 15.468(013) & 15.956(025) & 15.600(027) & $\cdots$ \\
701.57 & $\cdots$    & 15.910(023) & 15.563(021) & $\cdots$ \\
705.53 & 15.525(015) & 15.979(024) & 15.644(021) & $\cdots$ \\
710.61 & 15.628(014) & 16.098(022) & 15.805(026) & $\cdots$ \\
717.59 & 15.855(013) & 16.447(027) & 16.074(042) & $\cdots$ \\
719.60 & 15.906(021) & 16.582(033) & 16.152(047) & $\cdots$ \\
722.57 & 15.999(015) & 16.710(013) & 16.258(016) & 16.545(123) \\
723.53 & 16.026(020) & 16.814(235) & $\cdots$    & $\cdots$ \\
727.56 & 16.182(019) & 16.973(050) & 16.431(059) & $\cdots$ \\
731.54 & 16.322(019) & 17.166(059) & 16.599(063) & $\cdots$ \\
743.57 & 16.746(015) & 17.605(023) & 16.976(036) & $\cdots$ \\
756.55 & 17.184(027) & 17.998(052) & $\cdots$    & $\cdots$ \\
\enddata
\tablecomments{Values in parentheses are 1$\sigma$ measurement uncertainties in millimag.}
\end{deluxetable}

\end{document}